\documentstyle[a4p,12pt,epsfig,subfigure,array,cite,pennames]{article}
\newcommand{\Date}      {\rightline {18-July-1997}}
\newcommand{\PPEnum} {CERN-PPE/97-093}

\newcommand{\nZtoD}{0.1854}
\newcommand{\estatnZtoD}{0.0041}
\newcommand{\esysnZtoD}{0.0059}
\newcommand{\esysnZtoDBR}{0.0069}

\newcommand{\nZctoD}{1.041}
\newcommand{\estatZctoD}{0.020}
\newcommand{\esysZctoD}{0.040}

\newcommand{\nZbtoD}{1.334}
\newcommand{\estatZbtoD}{0.049}
\newcommand{\esysZbtoD}{0.078}

\newcommand{\xc}{0.515}
\newcommand{\estatxc}{0.002}
\newcommand{\esysxc}{0.009}

\newcommand{\nctoD}{0.222}
\newcommand{\estatctoD}{0.014}
\newcommand{\esysctoD}{0.014}

\newcommand{\nbtoD}{0.173}
\newcommand{\estatbtoD}{0.016}
\newcommand{\esysbtoD}{0.012}
\newcommand{\nRc}{0.180}
\newcommand{\estatRc}{0.011}
\newcommand{\esysRc}{0.012}
\newcommand{\ebrRc}{0.006}

\newcommand{\Ntol}{934}
\newcommand{\eNtol}{80}

\newcommand{\smcap}[1] {\caption[]{\small #1}}
\newcommand{\inmath}[1] {\ifmmode#1\else$#1$\fi}
\newcommand{\definmath}[2] {\def#1{\ifmmode#2\else$#2$\fi}}
\definmath{\GeV}  {{\mathrm{GeV}}}
\definmath{\GeVc} {{\mathrm{GeV}\!/c}}
\definmath{\GeVcc}   {{\mathrm{GeV}\!/c^2}}
\definmath{\MeV}  {{\mathrm{MeV}}}
\definmath{\MeVc} {{\mathrm{MeV}\!/c}}
\definmath{\MeVcc}   {{\mathrm{MeV}\!/c^2}}
\definmath{\MVm}  {{\mathrm{MV}\!/\mathrm{m}}}
\definmath{\keV}  {{\mathrm{keV}}}
\definmath{\keVcm}   {{\mathrm{keV}\!/\mathrm{cm}}}
\definmath{\kV}      {\mathrm{kV}}
\definmath{\km}      {\mathrm{km}}
\definmath{\meter}   {\mathrm{m}}
\definmath{\cm}      {\mathrm{cm}}
\definmath{\mm}      {\mathrm{mm}}
\definmath{\micron}  {\mu\mathrm{m}}
\definmath{\nm}      {\mathrm{nm}}
\definmath{\kg}      {\mathrm{kg}}
\definmath{\gram} {\mathrm{g}}
\definmath{\second}  {\mathrm{s}}
\definmath{\microsec}   {\mu\mathrm{s}}
\definmath{\degree}  {^\circ}
\definmath{\degC} {^\circ\mathrm{C}}
\definmath{\ohm}  {\Omega}
\definmath{\Mohm} {\mathrm{M}\Omega}
\definmath{\rad}  {\mathrm{rad}}
\definmath{\mrad} {\mathrm{mrad}}
\definmath{\nb}      {\mathrm{nb}}
\definmath{\dEdx} {{\mathrm d}E/{\mathrm d}x}
\newcommand{\gVf}  {g_{\mathrm{V}}^{\mathrm{q}}}
\newcommand{\gAf}  {g_{\mathrm{A}}^{\mathrm{q}}}

\def\Gamhad{{\rm \Gamma_{had}}}

\def\Gamcc {{\rm \Gamma_\cc}}

\def\cc{\ifmmode {{\mathrm c\bar{\mathrm c}}}
    \else {${\mathrm c\bar{\mathrm c}}$} \fi}
\def\bb{\ifmmode {{\mathrm b\bar{\mathrm b}}}
    \else {${\mathrm b\bar{\mathrm b}}$} \fi}
\def\qq{\ifmmode {{\mathrm q\bar{\mathrm q}}}
    \else {${\mathrm q\bar{\mathrm q}}$} \fi}
\definmath{\PWpm} {\mathrm{W}^{\pm}}      
\definmath{\Pgtp} {\tau^{+}}        
\definmath{\Pgtm} {\tau^{-}}        
\definmath{\Pgtpm}   {\tau^{\pm}}         
\definmath{\Pgn}  {\nu}          
\definmath{\Pagn} {\overline{\nu}}     
\definmath{\Pq}      {\mathrm{q}}
\definmath{\Paq}  {\overline{\mathrm{q}}}
\definmath{\Pf}      {\mathrm{f}}
\definmath{\Paf}  {\overline{\mathrm{f}}}
\definmath{\Pu}      {\mathrm{u}}
\definmath{\Pau}  {\overline{\mathrm{u}}}
\definmath{\Pd}      {\mathrm{d}}
\definmath{\Pad}  {\overline{\mathrm{d}}}
\definmath{\Ps}      {\mathrm{s}}
\definmath{\Pas}  {\overline{\mathrm{s}}}
\definmath{\Pc}      {\mathrm{c}}
\definmath{\Pac}  {\overline{\mathrm{c}}}
\definmath{\Pb}      {\mathrm{b}}
\definmath{\Pab}  {\overline{\mathrm{b}}}
\definmath{\Pt}      {\mathrm{t}}
\definmath{\Pat}  {\overline{\mathrm{t}}}
\definmath{\Pap}  {\overline{\mathrm{p}}}
\definmath{\Pan}  {\overline{\mathrm{n}}}
\definmath{\PaD}  {\overline{\mathrm{D}}}
\definmath{\PaDz} {\overline{\mathrm{D}}^{0}}
\definmath{\PaB}  {\overline{\mathrm{B}}}
\definmath{\PaBz} {\overline{\mathrm{B}}^{0}}
\definmath{\PsDpm}   {\mathrm{D}^{\pm}_{\mathrm{s}}}  
\definmath{\PcgLpm}  {\Lambda^{\pm}_{\mathrm{c}}}  
\definmath{\PcgL}  {\Lambda_{\mathrm{c}}}  
\definmath{\PsBz}  {\overline{\mathrm{B}^0_{\mathrm{s}}}} 
\definmath{\PbgL}  {\Lambda_{\mathrm{c}}}  
\definmath{\PD} {\mathrm{D}}     
\definmath{\PaDz} {\overline{\mathrm{D}}^{0}}
\definmath{\PDst} {{\mathrm{D}^{*}}}     
\definmath{\PDstp} {{\mathrm{D}^{*{\scriptscriptstyle +}}}}     
\definmath{\PDstm} {{\mathrm{D}^{*{\scriptscriptstyle -}}}}     
\definmath{\PDstpm} {{\mathrm{D}^{*{\scriptscriptstyle \pm}}}}     
\definmath{\PK}{{\mathrm{K}^-}}
\definmath{\Pgp}{{\pi^+}}
\definmath{\PaK}{{\mathrm{K}^+}}
\definmath{\Pagp}{{\pi^-}}

\definmath{\PgLz} {{\Lambda^{0}}}        
\def\D0{\ifmmode {{\mathrm D^0}} \else {${\mathrm D^0}$}\fi}
\def\Z0{\ifmmode {{\mathrm Z^0}} \else {${\mathrm Z^0}$}\fi}
\newcommand{\massof}[1] {m_{\smash{#1}\mathstrut}}

\newcommand{\mPZ} {\massof{\mathrm{Z}}}

\newcommand{\epem}   {\Pep\Pem}

\newcommand{\qqbar}  {\Pq\Paq}

\newcommand{\ccbar}  {\Pc\Pac}
\newcommand{\bbbar}  {\Pb\Pab}

\newcommand{\BzBzbar}   {\PBz\!\!-\!\PaBz}
\newcommand{\Gammaof}[1]   {\Gamma_{\!\smash{#1}\mathstrut}}

\newcommand{\Gcc}    {\Gammaof{\ccbar}}
\newcommand{\Gbb}    {\Gammaof{\bbbar}}
\newcommand{\Ghad}      {\Gammaof{\mathrm{had}}}
\newcommand{\Gqq}    {\Gammaof{\qqbar}}

\definmath{\BR}{\mathrm B}
\definmath{\NDP}{N_{\PDst\pi}}
\definmath{\NLP}{N_{\ell\pi}}
\definmath{\fctoD}{{\mathrm f}\,(\Pc\to\PDstp X)}
\definmath{\fbtoD}{{\mathrm f}\,(\Pb\to\PDstp X)}
\definmath{\fqtoD}{{\mathrm f}\,(\Pq\to\PDstp X)}

\newcommand {\downto}
         {\mbox{ \begin{picture}(14,10)
                    \put(0,10){\line(0,-1){5.0}}
                    \put(2,5){\oval(4,4)[bl]}
                    \put(1,0){\makebox(0,0)[bl]{$\rightarrow$}}
                 \end{picture} }}

\def\etal{et al.}

\begin{document}
%
%
\begin{titlepage}
%
%
\begin{center}
   \Large
    EUROPEAN LABORATORY FOR PARTICLE PHYSICS
\end{center}
\bigskip
\begin{flushright}
    \PPEnum\\
    \Date
\vskip 2cm 
\end{flushright}
%
%
\begin{center}
    \huge\bf\boldmath
Measurement of \\
${\rm f}(\Pc \to\PDstp X)$, ${\rm f}(\Pb \to\PDstp X)$  and
$\Gamma_{{\rm c\bar c}}/\Gamma_{\rm had}$ using $\PDstpm$ Mesons
\end{center}
\bigskip


\begin{center}
{\bf \large THE OPAL COLLABORATION}
\end{center}
\bigskip
%
%
\begin{abstract}

\noindent 
The production rates of $\PDstpm$ mesons in charm and bottom events 
at centre-of-mass energies of about $91~\GeV$
and the partial width of primary $\ccbar$ pairs 
in hadronic $\PZz$ decays 
have been
measured at LEP using almost 4.4 
million hadronic $\PZz$ decays
collected with the OPAL detector between 1990 and 
1995. 
Using a combination of several charm quark tagging methods based on fully and 
partially reconstructed $\PDstpm$ mesons, and a bottom 
tag based on identified muons and electrons, 
the hadronisation fractions of charm and bottom 
quarks into $\PDstpm$ mesons have been found to be
\begin{center}
\begin{tabular}{lcl}
${\fbtoD} = \nbtoD \pm \estatbtoD \pm \esysbtoD $  & and &
$ {\fctoD} = \nctoD \pm \estatctoD \pm \esysctoD \ .$ 
\end{tabular}
\end{center}
The fraction of $\ccbar$ events in hadronic $\PZz$ decays,
$\Gcc/\Ghad=\Gamma(\PZz\to\ccbar) / \Gamma(\PZz\to\mathrm hadrons)$,
is determined to be 
$$ \Gcc/\Ghad = \nRc \pm \estatRc \pm \esysRc \pm \ebrRc \ .$$
In all cases the first error is statistical, and the second one systematic.
The last error quoted for $\Gcc/\Ghad$ is due to external branching ratios. 
\end{abstract}
\vskip 1cm 
\medskip
\bigskip
\vspace{5cm}
\begin{center}
Submitted to Zeitschrift f\"ur Physik C
\end{center}
\end{titlepage}

\newpage

\begin{center}THE OPAL COLLABORATION 
\end{center}
\smallskip
\begin{center}{
K.\thinspace Ackerstaff$^{  8}$,
G.\thinspace Alexander$^{ 23}$,
J.\thinspace Allison$^{ 16}$,
N.\thinspace Altekamp$^{  5}$,
K.J.\thinspace Anderson$^{  9}$,
S.\thinspace Anderson$^{ 12}$,
S.\thinspace Arcelli$^{  2}$,
S.\thinspace Asai$^{ 24}$,
D.\thinspace Axen$^{ 29}$,
G.\thinspace Azuelos$^{ 18,  a}$
% ...
}\end{center}\bigskip
\bigskip
% ...
$^{  2}$Dipartimento di Fisica dell' Universit\`a di Bologna and INFN,
I-40126 Bologna, Italy
\newline
% ...
$^{  5}$Cavendish Laboratory, Cambridge CB3 0HE, UK
\newline
% ...
$^{  8}$CERN, European Organisation for Particle Physics,
CH-1211 Geneva 23, Switzerland
\newline
$^{  9}$Enrico Fermi Institute and Department of Physics,
University of Chicago, Chicago IL 60637, USA
\newline
% ...
$^{ 12}$Indiana University, Department of Physics,
Swain Hall West 117, Bloomington IN 47405, USA
\newline
% ...
$^{ 16}$Department of Physics, Schuster Laboratory, The University,
Manchester M13 9PL, UK
\newline
% ...
$^{ 18}$Laboratoire de Physique Nucl\'eaire, Universit\'e de Montr\'eal,
Montr\'eal, Quebec H3C 3J7, Canada
\newline
% ...
$^{ 23}$Department of Physics and Astronomy, Tel Aviv University,
Tel Aviv 69978, Israel
\newline
$^{ 24}$International Centre for Elementary Particle Physics and
Department of Physics, University of Tokyo, Tokyo 113, and
Kobe University, Kobe 657, Japan
\newline
% ...
$^{ 29}$University of British Columbia, Department of Physics,
Vancouver BC V6T 1Z1, Canada
\newline
% ...
$^{  a}$ and at TRIUMF, Vancouver, Canada V6T 2A3
\newline
% ...
\newline

\newpage

\section{Introduction}
\label{sec-intro}
The production of heavy quarks in the decay of the $\PZz$ boson
and their hadronisation have been the
subject of considerable interest over the last few years.
In particular the fractions with which
the $\PZz$ boson decays into quark pairs of flavour q have
been studied 
extensively in $\PZz\to\bbbar$ 
decays~\cite{bib-ALEPHRb,bib-DELPHIRb,bib-L3Rb,bib-OPALRb},
in $\PZz\to\ccbar$ decays~\cite{bib-ALEPHRc,bib-DELPHIRc,bib-OPALD*,bib-OPALcharm}
and in light flavour events~\cite{bib-OPALRuds}.
The fraction of $\bbbar$ events in $\PZz$ decays has been measured 
with very good precision.
To achieve this goal, very efficient and pure bottom tagging methods have been developed, 
resulting in samples of events that are nearly free of 
non-bottom backgrounds. Significantly fewer and less precise 
measurements exist of the equivalent quantity for $\ccbar$ events
or for light flavour events.  
In particular the selection of a pure $\ccbar$ sample has met with many
difficulties, and the efficiencies and purities 
achieved by charm tags are inferior to those for bottom tags. 
The reason for this is that charmed 
hadrons are lighter and shorter lived than 
bottom hadrons, and are similar enough to most light hadrons to make 
a separation very difficult. 
However, the precise knowledge of the partial widths for different 
flavours constitutes an important test of the 
predictions of the Standard Model,
since in lowest-order Born approximation 
the partial $\PZz$ decay width
to $\qqbar$, $\Gamma_{\qqbar}$,
is related to the coupling constants of the 
vector and axial vector current, $\gVf$ and $\gAf$: 
\begin{equation}
\Gamma_{\qqbar} = N_c^{\mathrm q}\;\frac{G_\mu \mPZ^3}{6 \pi \sqrt 2 } \bigl (
       (\gVf)^2 + (\gAf)^2 \bigr )  \ .
\end{equation}
Here  $G_\mu$ is the
Fermi decay constant and $\mPZ$ the \PZz\ mass.
The factor $N_c^{\mathrm q}=3$ denotes the number of colours. 
Higher order electroweak and QCD corrections to the 
$\PZz$ propagator and $\qqbar$ vertex that modify $\Gamma_{\qqbar}$
essentially cancel in the ratio $\Gamma_{\qqbar}/\Gamma_{\mathrm had}$ except 
in the case of $\PZz\to\bbbar$, where a small dependence on 
the Higgs mass and on the precise value of the top quark mass is introduced. 
The ratio $\Gamma_{\qqbar}/\Gamma_{\mathrm had}$ therefore is the preferred 
measurable quantity, for which precise predictions exist in the context 
of the Standard Model 
for the quark flavours u,d,s and c, almost independent of 
unknown quantities. 

In this paper a measurement of the fraction of primary $\ccbar$ pairs 
produced in the decays of $\PZz$ bosons is presented. At the same time 
the hadronisation fractions $\fctoD$ and 
$\fbtoD$ are measured. The analysis is 
based on the identification of charged $\PDstpm$ mesons, electrons and 
muons. 




The hadronisation fractions $\fctoD$ and 
$\fbtoD$ are measured using a double 
tagging technique. 
To determine $\fctoD$, charged $\PDstpm$ mesons are sought 
in both event hemispheres\footnote{The plane separating the 
two hemispheres in an event 
is defined perpendicular to the thrust axis of the event.}. 
The hadronisation fraction $\fbtoD$ is determined in events 
tagged by a hard lepton in one hemisphere and a $\PDstpm$ in the other 
hemisphere. 
Comparing the number of such double tagged events with the 
number of singly reconstructed $\PDstpm$ mesons or leptons, 
the hadronisation fractions can be extracted with minimal 
model dependence and without explicit knowledge of the $\PDstpm$ or 
lepton reconstruction efficiencies. 

The ratio of the charm partial width to the 
total hadronic width, $\Gcc/\Ghad$, is determined 
from the hadronisation fraction 
\mbox{$\fctoD$} and from a measurement 
of the total production rate of $\PDstpm$ mesons in $\PZz\to\ccbar$ 
events, $\Gcc/\Ghad \cdot \fctoD$. 
This rate is measured in this paper using 
a particularly well understood $\PDstpm$ decay mode, 
the decay\footnote{
Charge conjugation is assumed throughout this paper.}
$\PDstp\to\PDz\Pgp, \PDz\to\PK\Pgp$.

Both the measurement of the hadronisation fraction and 
the measurement of $\Gcc/\Ghad$ 
rely heavily on the reconstruction of 
$\PDstp$ mesons using two different techniques. 
Therefore the discussions in the first part of the 
paper concentrate on these technical aspects of the 
analysis.
In the first technique described in 
section~\ref{sec-ctag} the $\PDstp$ mesons are identified 
in a number of different decay channels by 
reconstructing all charged decay products. 
Since a significant 
contribution to the $\PDstp$ sample is from bottom hadron decays, 
a method has been developed to separate the different 
sources and is discussed in some detail. 
The second method of $\PDstp$ reconstruction is described 
next. It is a much 
more inclusive method, where only the pion in the decay 
$\PDstp\to\PDz\Pgp$ is used as the 
tag for the $\PDstp$. In the last part of the section 
the tagging of $\PZz\to\bbbar$ events using leptons is summarised. 

In the second part of the paper the different 
measurements are presented. In section~\ref{sec-ndstar}
the determination
of the total rate of $\PDstp$ production in hadronic $\PZz\to\ccbar$ 
decays is described. This is followed 
in section~\ref{sec-fctoD}
by a presentation of 
the double tagging technique used to measure 
the hadronisation fractions for both bottom and 
charm events.  Finally the results are combined to 
derive the relative partial width $\Gcc/\Ghad$.
The results reported in this paper supersede the ones 
given in~\cite{bib-OPALD*}, and complement the 
measurement of $\Gamma_{\ccbar}/\Gamma_{\mathrm had}$ reported
in~\cite{bib-OPALcharm}.




\section{Analysis Principle}
\label{sec-anal}
The main goal of this analysis is the measurement 
of the hadronisation fractions $\fctoD$ and $\fbtoD$ 
and of $\Gamma_{\ccbar}/\Gamma_{\mathrm had}$. 
A double tagging technique is used to minimise 
model dependencies. However, because charm tags are rather 
inefficient a full double tag, where the same tag is applied 
to both hemispheres of the event, cannot be used. Instead two 
different charm tags are applied, one, which 
is pure, but has a comparatively small efficiency; and the other, 
which is more efficient, but less pure. 
The general strategy for the measurement of the 
hadronisation fractions is that a charm or bottom enriched 
sample is selected by applying the high purity charm or bottom tag 
to one hemisphere of the event, and then searching 
for $\PDstp$ mesons in the opposite hemisphere using the 
more efficient, less pure tag. 
Neglecting for simplicity any background from other 
flavours, the number of events of flavour q is given by 
\begin{equation}
N_{\mathrm tag1} \sim {\Gqq \over \Ghad}\; {\fqtoD} \; \epsilon_{\mathrm tag1}\ ,
\label{eq-single}
\end{equation}
where $\epsilon_{\mathrm tag1}$ is the efficiency to select an event of 
flavour $\Pq$ using the pure tag. 
The number of events, where a $\PDstp$ mesons is simultaneously 
identified in the second hemisphere, is given by
\begin{eqnarray}
N_{\mathrm tag1\;tag2} &\sim& {\Gqq \over \Ghad}\;
     {\fqtoD}\;\epsilon_{\mathrm tag1}\times {\fqtoD}\;
\epsilon_{\mathrm tag2}\nonumber\\
     &=& N_{\mathrm tag1} \times {\fqtoD}\;\epsilon_{\mathrm tag2} \ .
\label{eq-double}
\end{eqnarray}
Here $\epsilon_{\mathrm tag2}$ 
is the efficiency to tag a $\PDstp$ mesons using the second, 
efficient, tag in the flavour tagged sample.
From the ratio of the number of double tagged events to the number 
of single tagged events
the hadronisation fraction $\fqtoD$ can be measured essentially without 
further assumptions or inputs. 
In this analysis the high purity 
flavour tags are an exclusive $\PDstp$ tag for 
$\PZz\to\ccbar$ events, 
and a lepton tag for $\PZz\to\bbbar$ events. The $\PDstp$ tag, 
applied to the sample of flavour tagged events,
is based on a very inclusive method of identifying 
$\PDstp$ mesons using only the pion from the decay 
$\PDstp\to\PDz\Pgp$.


Significant backgrounds however exist from other than the desired 
flavours. In addition the efficiency to find a $\PDstp$ meson in 
the flavour tagged sample is not independent of the 
flavour tag itself. 
Background is particularly important in the case 
where the flavour tag is the charm tag. A
significant part of the sample of $\PDstp$ mesons 
originates from $\PZz\to\bbbar$ events, and also
non-negligible contributions from combinatorial background events 
are found.
The charm tagged sample is
selected by fully reconstructing $\PDstp$ mesons that decay 
into a particular final state $\PK$Y with a
branching ratio ${\cal B}=\BR({\PDstp\to\PDz \Pgp}) \; \BR({\PDz \to \PK Y})$.
The number of such events, $N_{\PDstp}$, is given by
\begin{equation}
N_{\PDstp} = 2 N_{\rm had} \cdot
 \sum_{{\Pq=\Pb,\Pc}} \bigl({ \Gqq\over\Ghad } f_{\Pq}\;{\rm f\,}({\Pq\to\PDstp X}) 
       \; {\cal B}\; \epsilon^{\rm q}_{\PDstp}\bigr) + 
       N^{\mathrm bgd}.
\end{equation}                      
Here, $N_{\rm had}$ is the number of 
hadronic $\PZz$ decays used, 
$\Gqq/\Ghad$ is the relative partial width 
for a $\PZz$ to decay into a quark-antiquark pair of flavour $\Pq$, 
$f_{\Pq}$ is the fraction of events with flavour $\Pq$ in the sample,
$\epsilon^{\rm q}_{\PDstp}$ is the efficiency to reconstruct a 
$\PDstp$ meson in a ${\rm q}\to\PDstp$ hemisphere, 
and $N_{\mathrm bgd}$ is the number of background events in the sample. 
The fractions $f_{\Pq}$ satisfy the condition $f_{\Pc}+f_{\Pb}=1$.

%
In this sample of flavour tagged events $\PDstp$ mesons 
are sought in the opposite hemisphere using the inclusive 
$\PDstp$ reconstruction. 
Background in the sample 
is reduced by requiring that the two $\PDstp$ candidates have 
opposite charges. The contribution from $\Pb\to\PDstp$ 
decays
is further reduced since some of the bottom hadrons will 
have mixed before decaying into a $\PDstp$ meson.
The number of double tagged events is therefore given by
\begin{equation}
\NDP=(N_{\PDstp} - N^{\mathrm bgd})\cdot \Bigl [ 
         f_{\Pc}\; {\fctoD}\;\epsilon_{\PDst\pi}^{\Pc} + 
        f_{\Pb}\;(1-\chi_{\rm eff}) 
            \; {\fbtoD}\;\epsilon_{\PDst\pi}^{\Pb}\Bigr ] 
              \cdot {\cal B}_* + N^{\mathrm bgd}_{\PDst \pi}\ ,
\label{eq-doubletag}
\end{equation}
where ${\cal B}_* = \BR(\PDstp\to \PDz \Pgp)$, $\epsilon_{\PDst\pi}^{{\rm q}}$
is the efficiency for tagging a $\PDstp$ meson using the inclusive tagging
method in a $\PZz\to\qqbar$ event, when a $\PDst$ meson has been 
identified in the opposite hemisphere, and $ \chi_{\rm eff}$ is an 
effective 
$\BzBzbar$ mixing parameter applicable to the selected sample of events.
It is interesting to note that eq.~\ref{eq-doubletag} 
does not depend on the efficiency of the high purity flavour tag, 
but only on the efficiency of reconstructing 
a $\PDstp$ meson inclusively.

The number of double tagged events given in 
equation~\ref{eq-doubletag} is proportional to both $\fctoD$ and 
to $\fbtoD$. To separate the components the analysis is done twice, 
once as shown in eq.~\ref{eq-doubletag} in a charm enriched sample, tagged 
by the presence of $\PDstp$ mesons, and once in a bottom enriched
sample, selected through hard leptons. The latter analysis 
is mostly sensitive to $\fbtoD$, the former to $\fctoD$. 
By fitting the two samples simultaneously both hadronisation fractions 
are determined. 

The charm partial width, $\Gcc/\Ghad$, is determined from 
the measured hadronisation fraction \mbox{$\fctoD$} and from the total
rate with which $\PDstp$ mesons are produced 
in $\PZz\to\ccbar$ decays,
\mbox{$\Gamma_{\ccbar}/\Gamma_{\mathrm had} \; \fctoD$}.
This is measured using 
a particularly well understood $\PDstp$ decay mode, 
$\PDstp\to\PDz\Pgp, \PDz\to\PK\Pgp$, which facilitates 
the source separation and the background subtraction. 
The only quantity which is not measured in this 
analysis, but has to be taken from external sources, is 
the branching ratio $\BR(\PDstp\to\PDz\Pgp)$. 

\section{The OPAL Detector and Event Selection}
\label{sec-hadsel}
A detailed description of the OPAL detector can be found 
elsewhere \cite{bib-OPALdetector}.
This analysis relies heavily on the precise reconstruction of 
charged particle tracks and primary and secondary vertices
in the event. This is achieved using a 
combination of two layers of a  
high precision silicon micro-vertex detector, installed nearest to the 
primary interaction point, and a system of large-volume gas-filled 
drift chambers
which combine excellent spatial resolution with very 
good particle identification 
capabilities via the measurement of the specific energy loss of tracks.
The whole central tracking system is immersed 
in a magnetic field of $0.435$ T, oriented along the direction 
of the electron beam.
These central tracking detectors are 
surrounded by both electromagnetic and hadronic 
calorimeters with good energy 
resolution, providing nearly hermetic coverage over the full solid angle, 
and by a system of muon 
chambers on the outside of the detector.

Hadronic $\PZz$ decays are selected based on 
the number of reconstructed charged tracks and the energy deposited
in the calorimeter~\cite{bib-OPALmh}. The total hadronic event selection 
efficiency is found to be $(98.7\pm0.4)\%$. 
The selection slightly changes the flavour composition of 
the sample. This flavour bias is found to be 
less than $0.1\%$~\cite{bib-OPALRb}. 
The analysis uses an initial sample of $4 \thinspace374 \thinspace410$ 
hadronic 
decays of the $\PZz$ collected with the OPAL detector 
between 1990 and 1995. 

Jets are reconstructed in the events 
by the cone jet finder~\cite{bib-CONE} with a cone radius,
$R_{\mathrm cone}$, set 
to $0.7$, and a minimum cone energy 
of at least $5~\GeV$. 
Events are only accepted if at least two jets are reconstructed.
To ensure that the event is mostly contained 
in the sensitive detector volume, the absolute value of the 
cosine of the polar angle 
of the thrust axis with respect to the beam direction, 
$|\cos \theta_{\mathrm thrust}|$, has to be smaller than $0.9$. 

Tracks are used in the reconstruction if they pass 
loose track quality cuts requiring
$|d_0| < 0.5$ cm, 
$|z_0|  < 20 $ cm,
$p_{xy} > 0.250$ GeV and 
$n_{\rm CJ}>40$. 
Here $d_0$ is the distance of closest approach between the primary vertex 
and the track  measured in the 
plane perpendicular to the beam, $z_0$ the distance along the beam 
at this point, $p_{\rm xy}$ the momentum in the plane perpendicular 
to the beam, and $n_{\mathrm CJ}$ the number of hits on the track
which are reconstructed
in the main tracking chamber. 
The primary vertex in 
a collision is reconstructed from the charged tracks in the event
and constrained by the known 
average position and spread of the $\epem$ interaction point. 


Hadronic decays of the $\PZz$ have been simulated using the 
JETSET 7.4 Monte Carlo model  \cite{bib-JETSET} 
with parameters tuned to the data
\cite{bib-OPALtune}. A sample of approximately 10 million 
simulated events was available for this analysis. 
In all samples heavy quark fragmentation 
was implemented using the Peterson model \cite{bib-PETERSON}
with fragmentation parameters
determined from LEP data \cite{bib-LEPEW}. 
All samples have been  passed through a detailed
simulation of the OPAL detector \cite{bib-OPALGOPAL}
before being analysed using the same programs as for data. 




\section{Heavy Flavour Tagging Techniques}
\label{sec-ctag}
Three different tagging techniques are employed to identify 
$\PZz\to\ccbar$ and $\PZz\to\bbbar$ events. The charm tags are
based on the exclusive reconstruction of charged $\PDstp$
mesons (called ``exclusive tag'' in the following)
in five different decay chains, or on 
a more inclusive $\PDstp$ reconstruction
(called ``inclusive tag'' in the following). 
Bottom events are identified through the presence of an 
electron or a muon with large momentum and large transverse 
momentum relative to the direction of the jet containing the lepton. 
In this section the different tagging methods are described in 
some detail. Particular emphasis is placed on the method 
used to separate the contributions to the samples
tagged by the different methods, 
where large backgrounds are present, and on the 
systematic errors connected with this source separation method. 
Very similar techniques have been used in previous 
OPAL publications for charm tags~\cite{bib-OPALD*,bib-OPALD*AFB} 
and the bottom tag using leptons~\cite{bib-OPALlAFB,bib-OPALRb}.

\subsection{\boldmath The Exclusive Charm Tag}
\label{sec-dstar}
The exclusive charm tag is based on the reconstruction of 
charged $\PDstp$ mesons in five different decay channels: 
\begin{center}
\begin{tabbing}
  \hspace{5cm} \= \hspace{5cm} \= \kill
  \> $\PDstp \rightarrow \PDz\pi^+$ \\
  \> $\phantom{\PDstp \rightarrow }\hspace{4pt} 
            \downto {\rm K^-}\pi^+$\                     \> ``3-prong'' \\
  \> $\phantom{\PDstp \rightarrow }\hspace{4pt} 
            \downto {\rm K^-}{\rm e}^+ \nu_{{\rm e}}$\   \>``electron''\\
  \> $\phantom{\PDstp \rightarrow }\hspace{4pt} 
            \downto {\rm K^-}\mu^+\nu_{\mu}$\           \>``muon''\\ 
  \> $\phantom{\PDstp \rightarrow }\hspace{4pt} 
            \downto {\rm K^-}\pi^+\Pgpz$\                \> ``satellite'' \\
  \> $\phantom{\PDstp \rightarrow }\hspace{4pt} 
            \downto {\rm K^-}\pi^+\pi^-\pi^+$\           \>``5-prong''

\end{tabbing}
\end{center} 
In the following the electron and the muon channel are 
collectively referred to as ``semileptonic''.
No attempt is made to reconstruct the $\Pgpz$ in the satellite channel, or 
the neutrino in the two semileptonic channels. 
Electrons are identified based on their energy loss in the jet chamber 
and the energy deposition in the electromagnetic calorimeter. An 
artificial neural network trained on simulated events is used 
to perform the selection \cite{bib-OPALANN}. 
Muon candidates are identified by
associating tracks found in the central 
tracking system with
tracks in the outer 
muon chambers \cite{bib-OPALleptonID}.



In each channel a $\PDz$ candidate is formed by combining 
an appropriate number of tracks, corresponding to the 
number of charged decay products in the chosen decay mode,
assigning one to be a kaon, the rest to be pions or leptons, 
and calculating the invariant mass, $M_0$,  of the set of tracks. 
Candidates are selected if the tracks assigned to the 
decay products have the correct charges, and if the reconstructed mass lies 
within the expected range, defined by the mass resolution 
in the different channels. The exact values are given 
in table~\ref{tab-dstarcuts}.
After adding a further track as 
a candidate for the pion in the $\PDstp$ decay the 
combined mass, $M_*$,  is calculated and the candidate is selected 
if the mass difference $\Delta M = M_{*}-M_{0}$ 
is within given limits. Note that in cases where 
not all particles from the particular decay are reconstructed, 
the masses $M_0$ or $M_*$ do not correspond 
to the physical particle masses of the $\PDz$ or $\PDstp$ mesons, respectively.

For candidates with $x_{\PDstp}=E_{\PDstp}^{\mathrm cand}/E_{\rm beam}<0.5$
the particle identification power of the OPAL detector
is used to enrich the sample in true kaons. A probability $W_{\dEdx}^{\rm KK}$,
that the
difference between the measured specific energy loss, 
$\dEdx$, determined for a track of momentum $p$, 
and the $\dEdx$ value expected at that 
momentum for a kaon, is compatible with the kaon particle hypothesis, 
is calculated.
A candidate track has to have a probability of at least $2 \%$ 
to be accepted as a kaon candidate. To ensure a reliable $\dEdx$ 
measurement the number of charge measurements used in the 
$\dEdx$ calculation, $n_{\dEdx}$, has to be at least 20.

Background in the sample is further reduced by cutting on the helicity 
angle $\theta^*$, measured between the direction of the $\PDz$ candidate 
in the laboratory frame and the 
direction of the kaon in the rest frame of the $\PDz$ candidate. 
True kaons from $\PDz$ decays are 
expected to be isotropically distributed in $\cos \theta^*$, while 
background displays 
pronounced peaks at $\cos \theta^*=-1$ and,
particularly at low $x_{\PDstp}$,
at $\cos \theta^*=+1$. 

To avoid multiple counting of events if more than one $\PDstp$ 
candidate is found, only one candidate per event is accepted.
If more than one 
candidate is found per event a hierarchy is used to 
select the best one according to the 
signal purity.
A 3-prong decay is preferred over a 
semileptonic one, which in turn is preferred over a satellite, 
and a 5-prong is selected last. For the semileptonic 
channel an electron is preferred over a muon candidate. If more than one 
candidate is found within one channel, the one with 
$M_{0}$ closest to its nominal value of $M_{\PDz}=1.864~\GeV$ \cite{bib-PDG}
($1.6  ~\GeV$ for the satellite) is selected. 

\begin{figure}[p]
\begin{center}
\mbox{\epsfig{figure=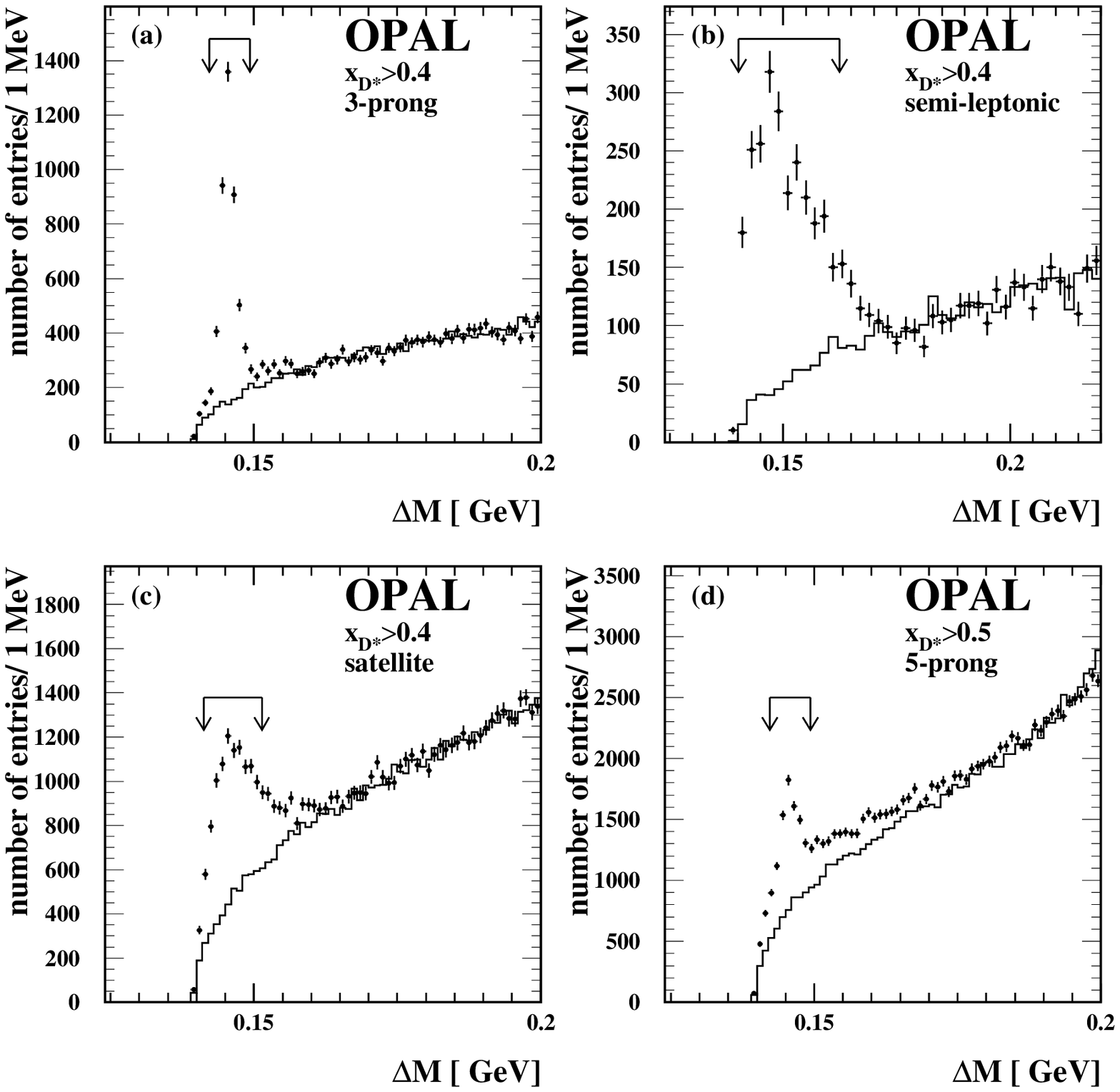,height=18cm}}
\end{center}
\smcap{\label{fig-delm} Distributions of the difference $\Delta M=M_{*}-M_{0}$
reconstructed in the four different $\PDstp$ channels.
The arrows indicate the selected signal region.  
(a) 3-prong decay mode, (b) the two semileptonic modes combined, 
(c) the satellite decay mode, and (d) the 5-prong decay mode. 
The points with error bars are the signal candidates.
Superimposed in each case (line histogram) are the 
background estimator distributions, normalised to the 
upper sidebands in $\Delta M$. }
\end{figure}

A detailed list of the cuts used in the different channels is 
given in 
table~\ref{tab-dstarcuts}.
\begin{table}[t]
\begin{center}
\begin{tabular}{|c|c||c|c|c|c|c|}
\hline
cut     & $x-$range & \multicolumn{2}{c|}{3-prong} & semileptonic  & satellite   & 5-prong   \\
\hline
\hline
$x_{\PDstp}$  &  & \multicolumn{2}{c|}{$0.4 $--$1.0$} & $0.4$--$1.0$& $0.4$--$1.0$  &  $0.5$--$1.0$        \\
\hline
$M_{0}$ [~\GeV]& full  &\multicolumn{2}{c|}{{$1.79$--$1.94$}}  & $1.20$--$1.80$   & $1.41$--$1.77$     & 
                    $1.79$-$1.94$       \\
\hline
$\Delta M$ [~\GeV]& full  &\multicolumn{2}{c|}{$0.142$--$0.149$}   & $0.140$--$0.162$ & $0.141$--$0.151$   & 
                    $0.142$--$0.149$        \\
\hline
$W_{\dEdx}^{\rm KK}$
    &$<0.5$  & \multicolumn{4}{c|}{$ >0.02$}&$-$\\
\hline
$n_{\dEdx}$  & $<0.5$ & \multicolumn{4}{c|}{20} & $-$ \\
\hline
$\cos \theta^*$ 
      & $<0.5$ & \multicolumn{4}{c|}{$-0.8 $--$ 0.8$}  &$-$ \\
      & $>0.5$ & \multicolumn{4}{c|}{$-0.9 $--$ 1.0$} & $-0.9$-$1.0$ \\
\hline
\end{tabular}
\end{center}
\smcap{\label{tab-dstarcuts}
List of selection cuts used in the \protect$\PDstp$ reconstruction. 
Note that both the scaled energy $x_{\PDstp}$ and the mass $M_0$
are effective quantities, calculated from the 
reconstructed tracks only. 
The exact meaning of the different quantities 
is explained in the text.}
\end{table}
The reconstructed invariant mass spectra in all channels 
exhibit characteristic peaks at $\Delta M = 0.145~\GeV$, close 
to the kinematic threshold of $\Delta M = m_{\pi}$. 
Clear signals are visible in all five channels, as shown
in figure~\ref{fig-delm}.

\subsubsection{Partially Reconstructed \protect\boldmath$\PDstp$ Mesons}
\label{sec-dstar-part}
A significant fraction of the sample of selected $\PDstp$ 
mesons are only partially reconstructed. This is particularly 
obvious for events where neutral decay products are not 
identified, as is the case for the satellite or the 
semileptonic channels. Even in decays where all decay products
are charged particles a fraction of events is present in the 
sample which are real $\PDstp$ decays, where however one 
or more tracks have been wrongly identified, or are missed 
completely. Such events are called ``partially reconstructed $\PDstp$ mesons''
if the slow pion of the $\PDstp\to\PDz\Pgp$ decay has been correctly found. 
These partially reconstructed $\PDstp$ mesons produce 
an enhancement in the $\Delta M$ spectrum very similar to the 
true signal. Only very few of these events are present 
in the 3-prong sample. They are much more important in the 
5-prong tagged events, where a clear tail is visible in 
the $\Delta M$ distribution for values above $0.145~\GeV$
(see figure~\ref{fig-delm}(d)).
Since such events originate from $\PDstp$ decays, they 
can still be used in the flavour tagging part of 
the analysis. 

\subsubsection{Combinatorial Background Estimation}
\label{sec-dstar-bgd}
The dominant background source 
is random combinations of tracks that
pass the applied cuts.
Only this combinatorial background component is considered
background for the flavour tagging, and a method 
has been developed to subtract only this component 
from the sample of tagged events. 
The combinatorial background component is described by an estimator
constructed entirely from data, optimised to exclude
partially reconstructed 
$\PDstp$ candidates. It is constructed using a hemisphere 
mixing technique first introduced in \cite{bib-OPALD*AFB}. The candidate 
for the pion in the $\PDstp\to\PDz\Pgp$ 
is taken from the opposite hemisphere relative 
to the rest of the candidate, 
and reflected through the origin, before being used 
in the calculation of the invariant mass. No 
requirements are placed on the charge of the $\PDstp$ background 
candidate, except that the total charge should be $\pm 1$. 
The resulting  distribution is used to define the shape of the background 
in $\Delta M$.
This method ensures that no true pions from the 
$\PDstp\to\PDz\Pgp$ decay are included in the 
estimator, and that the background shape does not 
exhibit any peak in the interesting $\Delta M$ region. 
The background distribution thus obtained is 
normalised to the candidate $\Delta M$ distribution in the range 
$0.18\;\GeV <\Delta M<0.20 \; \GeV$ ($0.19\;\GeV<\Delta M<0.22\;\GeV$ in the 
semileptonic channels).
Monte Carlo studies have shown that this ``reflected pion'' estimator 
reliably models the shape of the combinatorial background in the sample. 
The number of candidates and the estimated number of background 
events are given for all channels in table~\ref{tab-NDST}.
\begin{table}
\begin{center}
\begin{tabular}{|c|c|c|r@{$\pm$}l|}
\hline
\rule[-3mm]{0mm}{8mm} decay channel   &  $x_{\PDstp}$ range & 
$N_{\mathrm cand}$ & 
\multicolumn{2}{c|}{$N_{\mathrm bgd}$} \\
\hline
\hline
3-prong & $0.4-1.0$      & 4649 &  1034  & 28 \\
semileptonic & $0.4-1.0$ & 2485 &   587  & 23 \\
satellite & $0.4-1.0$    & 10086 &  4537 & 64  \\
5-prong & $0.5-1.0 $     & 9785 &   5208 & 64  \\
\hline
\rule[-3mm]{0mm}{8mm}total   && $27\thinspace 005$ & $11\thinspace 366$ & 107 \\
\hline
\end{tabular}
\smcap{\label{tab-NDST} Number of observed candidates in the signal region 
and the estimated number of these which are background events. 
The error quoted for the background is the statistical error of the background 
sample, and does not contain systematic effects. }
\end{center}
\end{table}

\subsubsection{Flavour Composition of the Exclusive \boldmath $\PDstpm$ Sample}
\label{sec-flavsep}
The $\PDstp$ mesons contained in the tagged sample originate
mostly in charm and bottom events, with a small contribution from events 
where a gluon splits into a pair of charm quarks. The latter 
is highly suppressed because of the high $x_{\PDstp}$ 
cut applied to the sample of selected events. Even more 
suppressed is the production of $\PDstp$ mesons from gluon 
splitting events into pairs of bottom quarks
because of the large 
mass of the bottom quark. 
In this section the determination 
of the composition of the sample is described. 

The composition of the tagged events 
is determined by applying three different bottom tags to the 
samples\footnote{The same method has been used in 
\cite{bib-OPALD*}, where additional 
details may be found.}, and combining the results. 
Assuming for simplicity that no background from light 
flavours is present in the sample, the number of events 
tagged by a particular bottom tag is given by 
%
\begin{equation}
{ N_{\Pb-\rm tag}  } = \bigl( f_{\Pb}\; {\cal P}_{\Pb}
                            + f_{\Pc}\; {\cal P}_{\Pc} \bigr) \; N_{\rm cand}\ .
\label{eq-bfrac}
\end{equation}
Here $f_{\Pb}=1-f_{\Pc}$ is the bottom (charm) fraction in the sample, 
and the ${\cal P}_{\Pb,\Pc}$ are the probabilities that an event
in which a $\PDstp$ meson has been identified 
is also tagged by the bottom tag. 
If these tagging probabilities are known, the bottom fraction can 
be calculated. The bottom tags are applied on a jet basis, 
both in  
the jet containing the exclusively reconstructed 
$\PDstp$ candidate (called the ``D-jet''), 
and in the remaining highest energetic jet in the event
(called the ``secondary jet'').

In practise additional backgrounds are present in the sample. 
The number of background events tagged with the 
different bottom tags and included in the sample is 
measured from data in independent background samples. 
For this a background sample for each tagging technique 
is prepared, the total number of events in the 
background sample is normalised to the number of 
background events measured, and the bottom 
tag is applied to this sample. 

The three different bottom tagging techniques 
are based on lifetime information, on jet-shapes and 
on hemisphere charge information. The first two have been 
used in earlier OPAL publications~\cite{bib-OPALD*,bib-OPALD*AFB}, 
and are only briefly reviewed. The last one will be covered 
in more detail.

Lifetime information is reconstructed in both jets used 
in this analysis. Vertices are reconstructed inclusively 
as in~\cite{bib-OPALgbb}, and a decay length significance
$d/\sigma$ is calculated, where $d$ is the distance between 
the primary and the secondary vertex, constrained by the 
jet direction, and $\sigma$ its error. Bottom 
events are identified by their large decay length significance 
values. The shape of the combinatorial background  
is estimated using the reflected pion 
technique discussed in section~\ref{sec-dstar-bgd}. The background estimator 
distribution is normalised to the sidebands in $\Delta M$, 
and is subtracted from the candidate distributions. 

Jet-shape information is used in the jet opposite 
the D-jet. The shapes are measured by a set of 
seven jet shape variables, which are defined in~\cite{bib-OPALD*}, 
and are combined using a neural net technique into one tag. 
The combinatorial background 
is estimated in data 
using a wrong charge technique, where background events 
are identified by the presence of a candidate with an 
unphysical charge combination of the decay products.

The third method uses the observation that the charge of the primary 
quark 
can be measured on a statistical basis using the hemisphere charge.
Since the correlations between the charges of the $\PDstp$ mesons and the 
sign of the charge of the primary quark are opposite 
for bottom and for charm quarks, measuring the $\PDstp$ charge and 
the primary quark charge in the opposite hemisphere 
provides some separation between bottom and 
charm events. 
The hemisphere charge 
is determined from all tracks in the hemisphere according to 
\begin{equation}
Q_{\mathrm hem} ={ { \sum_i |p_i|^\kappa q_i } \over  
{ \sum_i |p_i|^\kappa}} \ ,
\end{equation}
where $i$ runs over all tracks in a hemisphere, $p_i$ is the momentum 
component along the thrust axis
of track $i$ in the 
hemisphere, and $q_i$ is its charge. The exponent $\kappa$ is a weighting 
factor which 
has been optimised using Monte Carlo simulation to be $0.4$ for the purpose of 
flavour separation. 
Similar to the other 
two methods described a tagging efficiency $P_{\Pq}, \Pq=\Pb,\Pc$ 
is determined from Monte Carlo.
The shape of the background in the hemisphere charge is 
estimated from events tagged in sidebands of the $\Delta M$ 
distributions, $\Delta M >0.18~\GeV$, in the $\PDstp$ sample. 

\begin{figure}[p]
  \begin{center}
  \epsfig{file=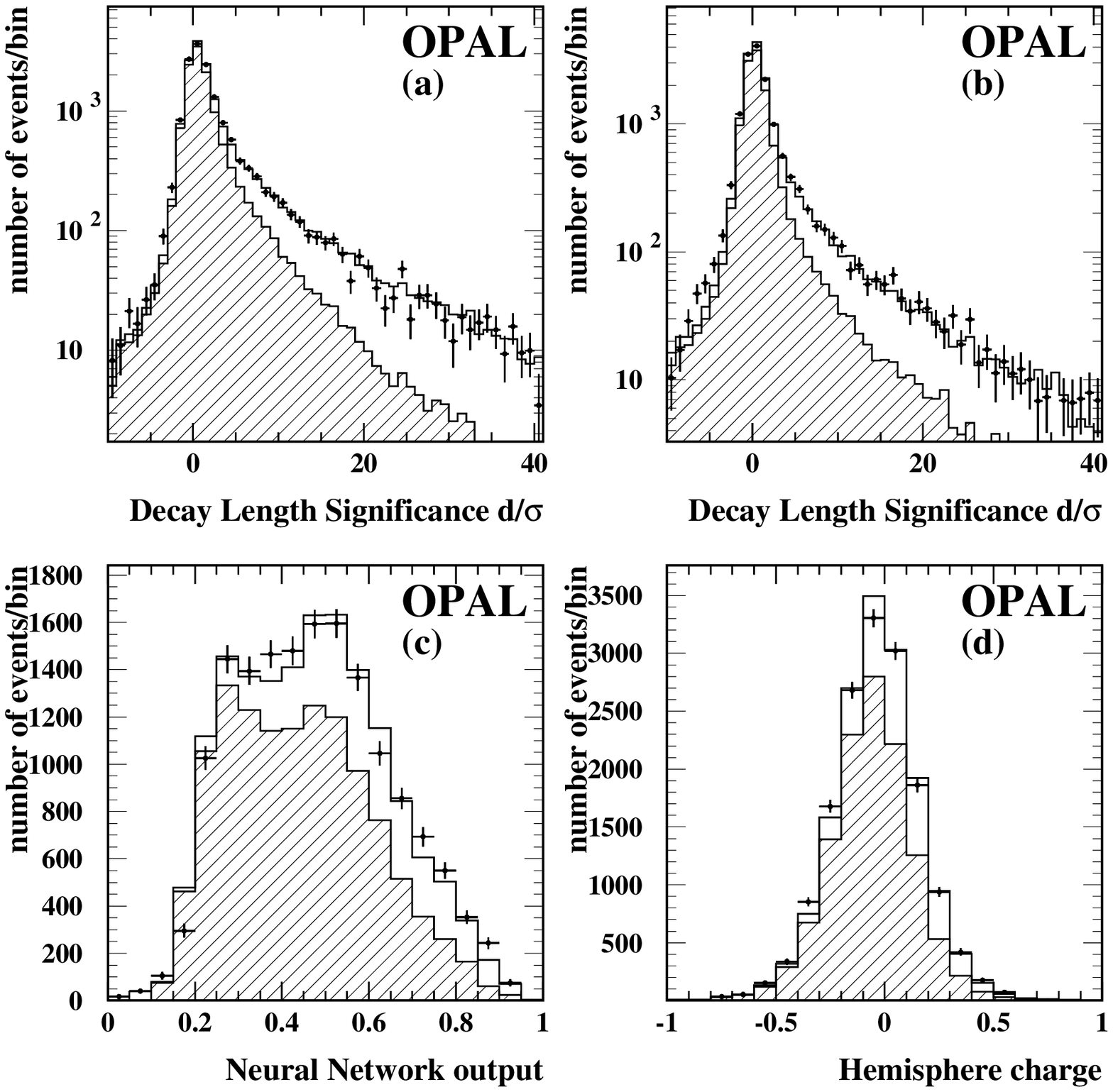,height=16cm}
  \end{center}
\smcap{\label{fig-tagvar} Distributions of the 
different tag-variables used in the flavour separation.
Shown are the data distribution after background subtraction (points with 
error bars), the equivalent Monte Carlo distribution for all candidates
(open histogram) and the predicted charm component (hatched histogram).
Shown are the 
(a) decay length significance in the jet with the exclusive $\PDstp$ candidate;
(b) decay length significance in the jet opposite the exclusive $\PDstp$ 
candidate; (c) distribution of the neural network based on jet-shape 
variables, and (d) distribution of the hemisphere charge.
}
\end{figure}

In all three cases the tagging probabilities are 
taken from the Monte Carlo simulation. 
The final fit for the flavour composition is performed 
simultaneously with the information from all three methods. 
It is done separately 
for each exclusive channel considered, and in bins of the scaled 
energy $x_{\PDstp}$ of the candidate. The most significant 
contribution to the separation comes from the decay length 
significance analysis, which contributes with a weight of 
$0.41$ from the $\PDstp$ hemisphere analysis, and
$0.27$ from the opposite hemisphere. 
The jet-shape analysis enters with a relative weight of $0.21$, 
the hemisphere charge with $0.11$.
Distributions of the tagging variables are shown 
in figure~\ref{fig-tagvar}.
Combining all exclusive channels 
the charm fraction in the $\PDstp$ tagged sample, for 
the range of $x_{\PDstp}$ detailed in table~\ref{tab-dstarcuts},
is determined to be 
\begin{equation}
  f_{\Pc}^{\PDstp} = 0.774 \pm 0.008\ ,
\end{equation}
where the error quoted is purely statistical. 

\subsubsection{Contribution from \boldmath ${\rm g}\to\ccbar$}
\label{sec-gcc}
Small contributions to the signal are expected from the splitting
of a gluon into a pair of charm quarks. This rate has been 
measured in~\cite{bib-OPALD*,bib-OPALgcc}, where the multiplicity 
of $\ccbar$ pairs produced in hadronic decays of the $\PZz$ is found to be 
$\bar n_{{\rm g} \to \ccbar} = 0.0238\pm 0.0048$. 
Using Monte Carlo simulation the fraction of events from this source 
in the selected sample of events 
is estimated and, after normalising to the measured rate, 
subtracted from the tagged sample of $\PDstp$ events. 
The total contribution of these events to the 
exclusively tagged sample is found to be $(0.2 \pm 0.1) \%$, 
where the error quoted is only due to 
Monte Carlo statistics. 

\subsubsection{Systematic Errors of the Flavour Separation Method}
\label{sec-flavsep-sys}

A number of errors are introduced by the flavour separation method 
employed. The errors quoted are relative errors 
on the charm fraction in the sample. 
\begin{itemize}
%
%
\item Detector resolution: The influence of the detector 
resolution on the tagging variables is studied in Monte Carlo 
by varying the resolutions in the tracking system by $\pm 10\%$ relative to 
values that optimally describe the data. The analysis is 
repeated and efficiencies are recalculated. The error quoted is the largest 
observed deviation in this study, and amounts to $0.6\%$.
\item Background modelling: The distributions in the tagging variables 
for background events are taken from estimator distributions 
determined using data. Possible differences between the estimator 
and the true background distributions are studied using 
Monte Carlo simulations. The separation is repeated with the 
background distribution taken from Monte Carlo, 
and the differences are used as systematic errors. 
Similarly, errors in the determination of the normalisation of background 
in the sample will change the reconstructed b-fraction (and, since they 
are totally anti-correlated, the $\Pc$ fraction). 
This has been studied by varying the background within 
its total error.
In total the error from these sources amounts to $1.7\%$.
\item Detector response: Possible inhomogeneities of the 
detector response as a function of $\cos \theta$ are studied 
by repeating the flavour separation in bins of $\cos \theta$, 
and by comparing the results with the overall determination. 
No significant differences are found.
\item Hemisphere correlations: Part of the  flavour separation is done 
in the hemisphere opposite to the reconstructed $\PDstp$ 
mesons. Small correlations are expected to exist between the two
hemispheres, which possibly might bias the measurement 
of the flavour composition. In the 
flavour separation these biases are taken into account 
by calculating the tagging efficiencies in bins of $x_{\PDstp}$ 
as measured from the exclusive D candidate. The size of the bias 
is estimated by recalculating the tagging efficiencies 
in only one bin of $0.4<x_{\PDstp}<1.0$, and repeating the analysis. 
This is done for all three tagging algorithms.
The resulting error is $0.4\%$.
\item Charm modelling: The jet-shape analysis 
is sensitive to the modelling of the response to charm 
events, which is taken from Monte Carlo simulation.
Possible modelling problems are investigated by comparing the 
network output distribution in an unbiased sample of 
hadronic $\PZz$ decays in data and Monte Carlo. All observed differences
are assumed to come from charm modelling problems, and 
a systematic error of $1\%$ is calculated. The same
method was used in~\cite{bib-OPALD*}.
\item Charm and bottom multiplicity: The vertex finder employed
is sensitive to the charged multiplicity from charm and bottom 
hadron decays in the sample. The multiplicity for heavy flavour 
decays in the Monte Carlo has been varied by reweighting simulated 
events, corresponding to the current experimental bounds of 
$\pm 0.2$ tracks for charm decays, and $\pm 0.35$ tracks for bottom 
decays~\cite{bib-LEPEW}.
Similarly the hemisphere charge technique is sensitive 
to the multiplicity, and its error is estimated 
using the same procedure. 
Overall this results in an error of $0.6\%$ for charm and 
$0.5\%$ for bottom. 
\item B hadron lifetime: The B hadron lifetime has been varied 
within its current experimental limits: 
In the D hemisphere the lifetimes of the different B species 
have been varied independently by 
$\pm 0.07 $~ps for the $\PBp$, $\pm 0.08$~ps for the $\PBz$, 
and $\pm 0.12$~ps for the $\PB_{\Ps}$ \cite{bib-PDG}. 
In the hemisphere opposite to the D meson, the mean B hadron lifetime
of \mbox{$(1.549\pm0.020){\mathrm ps}$~\cite{bib-PDG}} 
has been used and changed within its error. 
The total error is found to be $0.7\%$.
\item Charm lifetime: The lifetimes of the weakly-decaying charmed 
hadrons $\PDz$ and $\PDp$ has been varied independently 
by $\pm 0.004$ ps for the $\PDz$, and $\pm 0.015$ ps for the 
$\PDp$ \cite{bib-PDG}, corresponding to a total 
error of $0.4\%$.
\item $\BzBzbar$ mixing: Mixing of neutral B mesons changes 
the correlation between the charge of the quarks in the 
two hemispheres of the event. The hemisphere charge method 
is sensitive to the assumed value of the mixing through the 
bottom tagging efficiency. Two different mixing parameters 
have to be considered: the average B mixing in the hemisphere 
opposite the tagged $\PDstp$, and the effective mixing 
in tagged $\PDstp$ events. The latter is different because
the mixture of B mesons in $\PDstp$ tagged events 
is different from an unbiased sample. The derivation of this 
effective mixing is described later in this paper in 
section~\ref{sec-Bmix}, and is determined with an 
error of $\pm 17\%$. Compared to this the 
error on the average mixing is very small and is neglected. 
Applied to the b/c separation this translates into an 
error of $0.3\%$. 
\item Fragmentation model: The fragmentation functions 
used in Monte Carlo simulation for heavy flavour events influence 
all three tagging algorithms.
The Peterson \cite{bib-PETERSON} scheme is used for bottom 
and charm hadrons. To estimate the influence on the results the 
mean scaled energy of primary charm and bottom hadrons has been 
varied around their measured values of 
$\langle x_{\Pb}  \rangle= 0.702 \pm 0.008$ and 
$\langle x_{\PDstp} \rangle = 0.510 \pm 0.009$,
as suggested in~\cite{bib-LEPEW}, and the separation has 
been redone. 
This results in an error of $0.6\%$.
\end{itemize}
The final charm fraction in the $\PDstp$ tagged sample for 
the selected $x_{\PDstp}$ range is found to be  
\begin{equation}
  f_{\Pc}^{\PDstp} = 0.774 \pm 0.008 \pm 0.022\ ,
\end{equation}
where the first error is statistical, the second one systematic.
%

\subsection{The Inclusive Charm Tag}
\label{sec-slowpion}
The second, more inclusive, charm tag relies on the very special kinematical 
properties of the decay \mbox{$\PDstp \to \PDz \Pgp $}. Because of the small 
mass difference of only 145~\MeV~between the $\PDstp$ and the $\PDz$ 
very little phase space is left for the pion. In the laboratory 
frame this pion, called the ``slow pion'' in the following, 
is emitted essentially 
in the direction of the $\PDstp$ meson, with a maximal transverse
momentum $p_t$ relative to the $\PDstp$ direction of flight of $39 ~\MeV$. 
A charm tag is constructed from this by looking for an 
enhancement in the density of tracks along the $\PDstp$ flight 
direction. 

The fraction of slow pions from $\ccbar$ events is enhanced by requiring:
\begin{itemize}
\item $1.0~{~\GeV} < p_{\pi} < 3.0~{~\GeV}$\ ,
\end{itemize}
where $ p_{\pi}$ is the momentum of the pion candidate.
Kaon and electron contamination in the slow pion candidate sample
is reduced by using the particle identification 
power of the drift chamber in the OPAL detector, requiring
\begin{itemize}
\item $W^{\pi\pi}_{\rm dE/dx}~>~0.02$\  ,
\item $n_{\dEdx} > 20$\ ,
\end{itemize}
where $ W^{\pi\pi}_{\rm dE/dx}$ is the $\dEdx$ probability for a pion, 
and $n_{\dEdx}$ the corresponding number of measurements used,
as defined in section~\ref{sec-dstar}.

The flight direction of the $\PDstp$ meson is reconstructed inclusively 
by an iterative procedure. It uses the fact that decay products from 
heavy mesons are on average harder and more collimated 
than those from fragmentation 
tracks, leading on average to higher values of the 
rapidity\footnote{The momentum $p_z$
is measured relative to the jet axis. Charged particles are 
assumed to be pions, neutral particles photons.} 
$y=({1/2})\log [(E+p_z)/(E-p_z)]$
with respect to 
flight direction of the $\PDstp$ meson. 
The decay products are selected by first grouping all 
tracks and unassociated neutral clusters
in the event into jets using the cone algorithm, described in 
section~\ref{sec-hadsel}.
The jet axis is computed 
from all particles in the jet after removing the slow pion candidate itself. 
If tracks or clusters exist which have a rapidity 
measured relative to the direction of the jet, of less than $2.5$, 
the one with the smallest rapidity
is removed from the calculation, and the direction is recomputed. 
This procedure is repeated until all particles have a rapidity value 
above $2.5$, 
or the number of tracks or clusters is less than two. In this case 
the original jet direction is used. The direction determined in this manner
is used as an estimate of the $\PDstp$ flight direction. 
The resolution, measured as the width of the $p_t^2$ distribution 
at $50\%$ of its maximal value, is found to be $\sigma(p_t)=0.056\;\GeV$.
The efficiency with which $\PDstp\to\PDz\Pgp$ decays are selected using this 
method is around $40\%$ in $\ccbar$ events, and around 
$20\%$ in $\bbbar$ events.
A similar procedure was first introduced in~\cite{bib-DELPHIvcb}.


\subsection{\boldmath The Bottom Tag} 
\label{sec-lepttag}
A pure sample of bottom events is selected using a lepton tag. This 
sample will be used in the measurement of $\fbtoD$ as the 
flavour tagged sample, and takes the place of 
the exclusive $\PDstp$ tag described earlier. 
Electrons and muons are identified as described in section~\ref{sec-dstar}.
The sample is purified by requiring that electrons have 
a momentum larger than $2.0\; \GeV$, and a transverse momentum 
relative to the jet direction 
larger than $1.1\;\GeV$. Electrons are only reconstructed in the 
central region of the detector, if the polar angle is below 
$|\cos \theta| < 0.715$. 
Muons have to have a momentum above $3.0\;\GeV$, a transverse 
momentum larger than $1.2 \; \GeV$ and $|\cos \theta|<0.9$.
According to the Monte Carlo simulation this sample has a bottom purity of 
$(89.90 \pm 0.14) \%$. Of the remaining events $33\%$ are from 
semileptonic charm decays, and $66\%$ are misidentified leptons. 
 
\subsubsection{Systematic Errors of the Lepton Tag}
The purity of the lepton identification
is taken from Monte Carlo simulation. The following systematic errors 
have been investigated: 
\begin{itemize}
\item Detector resolution: The resolution 
of the tracking part of the detector is varied by 
$10\%$, resulting in an error on the purity of $0.1\%$.
\item Bottom fraction in $\PZz$ decays: The fraction of $\PZz\to\bbbar$ 
events in the Monte Carlo is re-weighted to the one 
measured by OPAL: $\Gamma_{\bbbar}/\Gamma_{\mathrm had}=0.2175 \pm 0.0022$
\cite{bib-OPALRb}. 
The error on the measurement is used to calculate the 
corresponding systematic error of $1.0 \%$.
\item Heavy flavour fragmentation: The fragmentation 
parameters in the Monte Carlo have been varied to change 
the mean scaled energy of charm and bottom hadrons around 
their experimental values of $0.702 \pm 0.008$ and $ 0.510\pm 0.009$ 
respectively. 
The error on the bottom purity is $0.8\%$.
\item Decay modelling: The momentum distribution of the lepton 
produced in a bottom or a charm decay influences the 
tagging efficiency. Following the recommendations in~\cite{bib-LEPEW} 
this was studied by reweighting the distribution in the Monte 
Carlo to different models. Models used are ACCM, ISGW and 
ISGW$^{**}$. The largest observed deviation is used as a systematic 
error, resulting in an error of $1.4\%$.
\item Semileptonic branching ratios: The semileptonic branching 
ratios $\BR(\Pb\to\ell)$ and $\BR(\Pc\to\ell)$ have been 
measured at LEP. The spectra in B-decay are determined also 
at lower energy machines. The values recommended 
in ~\cite{bib-LEPEW} are used, and the Monte Carlo is 
re-weighted to these measured values. Systematic errors 
are derived from the errors on the branching ratios. The 
error on the purity is $0.3\%$.
\item Hadronic background: Around $6\%$ of the sample 
of tagged leptons are hadrons, which were misidentified. In~\cite{bib-OPALRb},
the error of the mistagging rate has 
been determined to be $9.3\%$ in the electron sample, 
and $9.0\%$ in the muon sample. This translates into an 
error of $0.6\%$ of the bottom purity.
\end{itemize}
The total systematic error of the bottom purity in 
lepton tagged events is
found to be $2.1\%$. Note that the knowledge of 
the lepton reconstruction efficiency is not required in this analysis, as 
discussed in section~\ref{sec-anal}.

\section{\boldmath Production of $\PDstp$ Mesons in $\PZz\to\ccbar$ 
and $\PZz\to\bbbar$ Decays}
\label{sec-ndstar}
In this part the total production rates 
$\Gcc/\Ghad \cdot {\fctoD}\; 
            \BR(\PDstp\to\PDz\Pgp)\;\BR(\PDz\to\PK\Pgp)$ and
$\Gbb/\Ghad \cdot {\fbtoD}\; 
            \BR(\PDstp\to\PDz\Pgp)\;\BR(\PDz\to\PK\Pgp)$ are
determined. The results for $\fctoD$ is used later in the analysis to
determine 
the relative partial width $\Gcc/\Ghad$.
In addition the mean scaled energy of $\PDstp$ mesons in 
$\PZz\to\ccbar$ events, $\langle x_{\PDstp} \rangle_{\Pc}$
is measured, and the total 
multiplicity of charged $\PDstp$ mesons in $\PZz$ decays is 
given. 

The analysis is performed using the sample of fully 
reconstructed $\PDstp$ mesons in the 3-prong decay mode.
To minimise the number of $\PDstp$ mesons not observed
the $x_{\PDstp}$-range for this part of the 
analysis has been extended to 
$x_{\PDstp}=0.2$. The reconstruction method is similar to the 
one described in the previous section for the exclusive charm tag. 
An important difference however is the treatment of 
the partially reconstructed $\PDstp$ decays. While previously 
they have been treated as signal for the purpose of 
tagging the primary event flavour, they are background 
for the determination of the total rate of
$\PDstp$ meson production in this particular channel. 
Therefore the background subtraction is rediscussed in some 
detail, and a method of treating the contribution from 
such partially reconstructed decays is introduced. 

\subsection{Background Subtraction}
\label{sec-excl-bgd}
Backgrounds for the purpose of the measurement of the 
production rate of $\PDstp$ mesons are combinatorial 
background and partially reconstructed decays of $\PDstp$ mesons. 
The former is determined with essentially the 
same procedure as described above, except that the 
combinatorial background, after having been normalised, 
is not simply subtracted, but is fitted using a 
simple parametrisation. The contribution from 
partially reconstructed 
$\PDstp$ decays is measured in the data with a special procedure. 
In this part of the analysis 
no requirement is made that only one candidate be found in the channel,
unlike that for the sample of $\PDstp$ mesons used in 
the exclusive charm tag.  
This different treatment is possible since the 3-prong 
decay is very clean, and the number of partially 
reconstructed events is small. 

\begin{figure}[t]
  \begin{center}
  \begin{tabular}{cc}
    \epsfig{figure=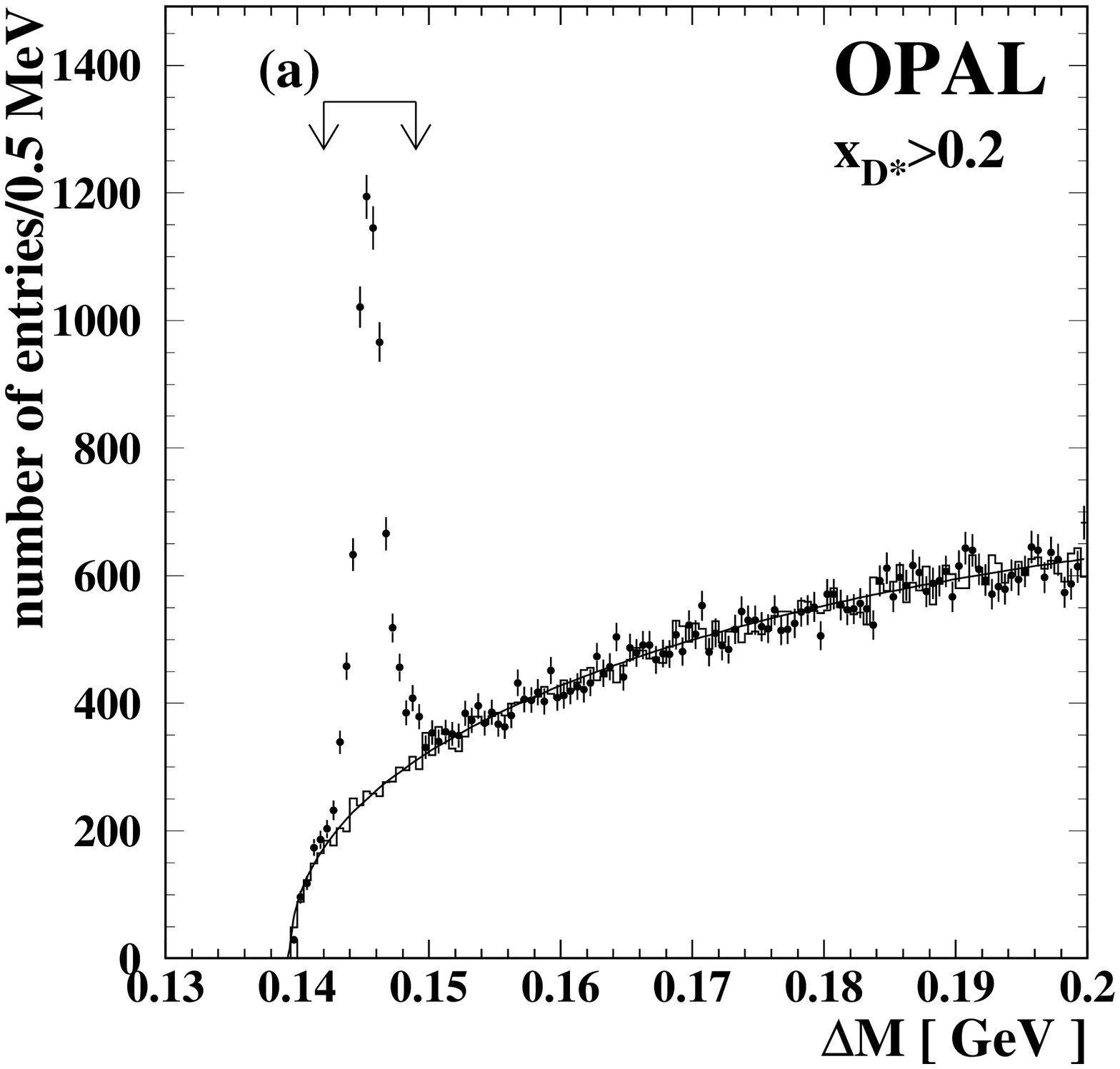,bbllx=0.5cm,bblly=1cm,bburx=22.5cm,bbury=19cm,height=8cm} & 
    \epsfig{figure=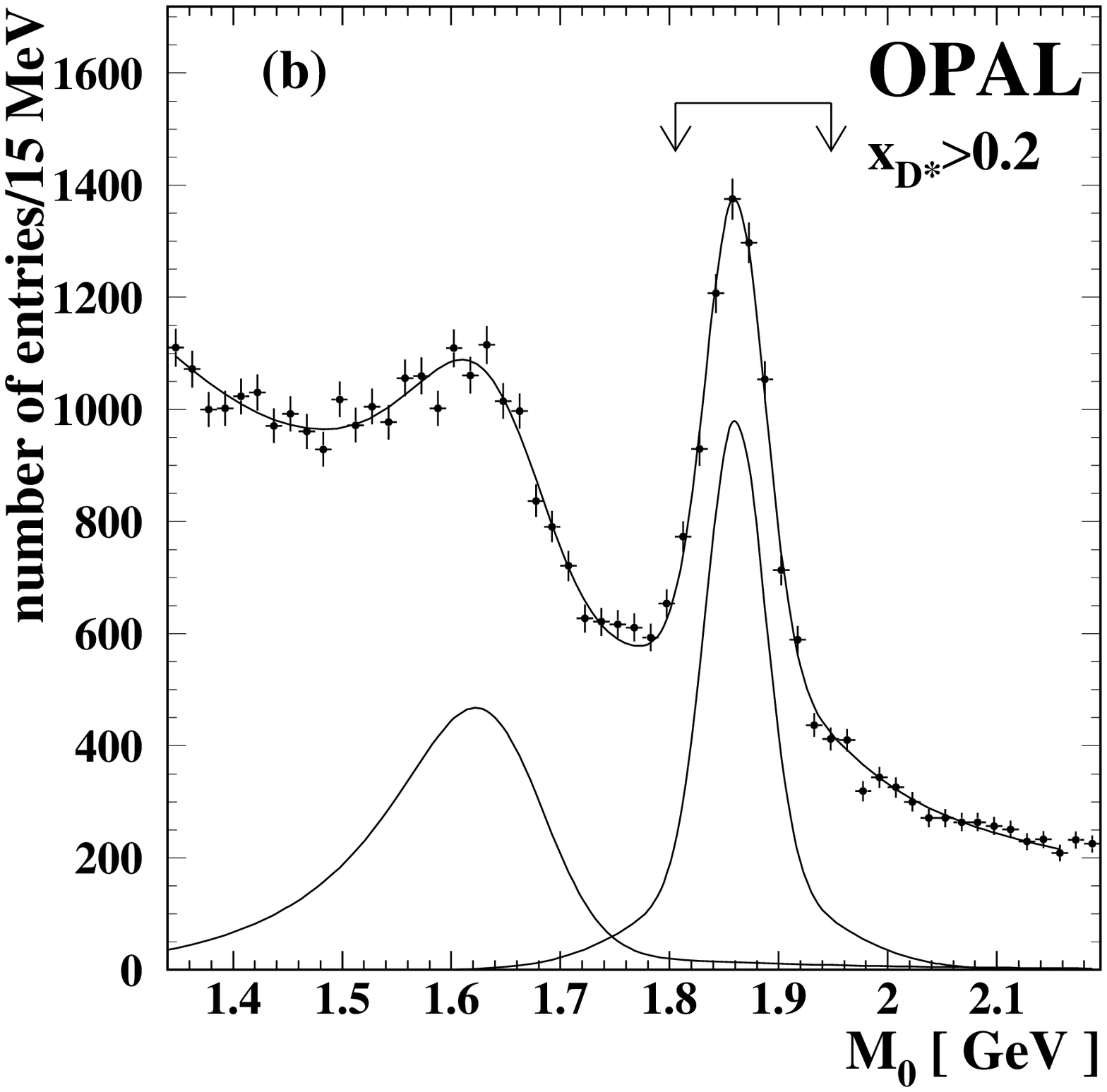,bbllx=4.5cm,bblly=1cm,bburx=26.5cm,bbury=19cm,height=8cm} 
  \end{tabular}
  \smcap{\label{fig-DM3pr} (a) 
Distribution of the mass difference $\Delta M=M_{*}-M_{0}$
reconstructed in the decay \protect$\PDstp\to\PDz\Pgp$, 
\protect$\PDz\to\PK\Pgp$. 
Superimposed is the background distribution obtained from the 
background estimator discussed in the text, and the result of the 
fit to this background estimator. The arrows indicate 
the selected signal region. (b) $M_{0}$ spectrum 
of $\PDstp$ candidates, with $M_0$ cut removed, and an 
additional $\Delta M$ cut applied. 
Shown are the data (points with error bars), 
the result of the fit as described in the text, and the two 
components from satellite and from fully reconstructed 
$\PDstp$ mesons, as obtained in the 
fit. The arrows indicate the selected signal region. 
}
  \end{center}
\end{figure}

The combinatorial background in the sample is determined 
as before from the reflected pion estimator. 
The $\Delta M$ distributions obtained using this estimator 
are normalised to the candidate $\Delta M$ distribution for 
$0.16\;\GeV < \Delta M < 0.2\;\GeV$. It is then parametrised using an 
empirical functional form 
\begin{equation}
f(\Delta M) = A \;
  ( { 1 \over m_{\Pgp}} ( \Delta M + B \; (\Delta M)^2 ))^C \ ,
\end{equation}
with $A,B$ and $C$ free parameters determined in the fit described below. 
The number of background events is determined in $16$ bins 
of $x_{\PDstp}$ between $x_{\PDstp}=0.2$ and $1.0$, in the 
signal region $0.142~\GeV < \Delta M < 0.149 ~\GeV$. The mass difference 
distribution for the 3-prong sample only, with the result of the background 
fit superimposed, is shown in figure~\ref{fig-DM3pr}(a) for the 
$x_{\PDstp}$ range between $0.2$ and $1.0$.

From Monte Carlo simulation about $8\%$ of the signal, 
after subtracting the combinatorial 
background, 
actually come from partially reconstructed $\PDstp$ mesons. 
They are mostly 
products of the following decays:
$\PDz\to\PK\Pgp\Pgpz$ ($4.3\%$),
$\PDz\to\PK\rm K^+$ ($1.8 \%$),
$\PDz\to\pi^-\Pgp$ ($0.7\%$),
$\PDz\to\pi^-\Pgp\Pgpz$ ($0.5\%$) and
$\PDz\to\PK\ell^+\nu$ ($0.3\%$). 
The numbers in brackets indicate 
the predictions from the Monte Carlo simulation for the contribution to 
the full $\PDstp$ sample from each source. 

Instead of using the Monte Carlo predictions the total 
contribution is measured in data 
in a simultaneous fit to the 
$M_{0}$ and the $\Delta M$ distributions of all candidate $\PDstp$ mesons. 
The $M_{0}$ distribution is examined for candidates where 
the mass difference $\Delta M$ is inside the tight signal 
region of $0.142\; \GeV< \Delta M < 0.149\;\GeV$, and 
no $M_0$ cut has been applied. 
Contributions from partially reconstructed $\PDstp$ mesons
are in general characterised by a peak in the $\Delta M$ 
distribution at the position expected for correctly identified 
$\PDstp$ mesons, but no peak-like structure in the 
$M_0$ distribution at the position expected for true $\PDz$ mesons. 
Therefore the difference between the number of reconstructed 
$\PDstp$ mesons as derived from the $\Delta M$ distribution 
and and from the $M_0$ 
distribution can be used as a measure of the fraction of 
partially reconstructed decays in the sample. 
A slight complication arises from the decay $\PDz\to\PK\PaK$, 
which is expected to peak just below the nominal $\PDz$ mass. Monte 
Carlo simulation is used to account for this. 

The number of $\PDstp$ candidates is extracted from the $M_0$ 
mass spectrum using a fit. The combinatorial background
is parametrised by an exponential 
function. The signal function has two contributions: 
the first describes the true 3-prong candidates, and is 
constructed from two gaussian functions, motivated from 
Monte Carlo simulation studies, 
with the second having three times the width and the same mean 
as the first. The second function parametrises the contribution 
from partially reconstructed decays in the signal region. It is 
dominated by satellite decays 
which cluster below the expected $\PDz$ mass of $1.865~\GeV$. 
As expected from the kinematics of this decay 
the mass distribution is approximately gaussian, with a 
significant tail towards smaller masses. It is parametrised by 
a gaussian convoluted with an exponential function. 
The decay constant in the exponential 
is fixed relative to the width of the gaussian function to the 
value obtained in the Monte Carlo. The other decays contributing 
to the partially reconstructed sample are described by an 
additional exponential function, added to the parametrisation 
of the satellite decay. This essentially adds a tail to the 
satellite function, which extends into the nominal 
$\PDz$ mass region. 
Monte Carlo is used to estimate the contribution coming 
from the $\PDz\to\PK{\mathrm K^+}$ decay, which is  not described by the tail. 
The shapes of the different fit functions have been tested in 
Monte Carlo simulated events, and are found to provide a 
good description of the mass spectra. 
Because of the complicated fit function, and because the 
fraction of partially reconstructed decays varies only 
slowly with $x_{\PDstp}$,
this fit is done in 
four equal-sized bins of $x_{\PDstp}$ between $0.2$ and $1.0$, 
instead of the 16 used in the $\Delta M$ fits. 
The number of $\PDstp$ mesons is obtained by 
integrating the signal function over the mass window 
$1.79\;\GeV < M_{0} < 1.94\;\GeV$. The fraction of partially reconstructed 
$\PDstp$ 
events in this bin is then calculated from the difference between the 
number of signal events determined with this fit, 
and the sum of the number of signal candidates over the 
appropriate $x_{\PDstp}$ bins found in the fit using the 
$\Delta M$ method. The $M_{0}$ spectrum for all candidates, 
with the results of the fit superimposed, is shown in 
figure~\ref{fig-DM3pr}(b).

Monte Carlo studies indicate that this method reliably reproduces 
the number of partially reconstructed $\PDstp$ mesons 
in the sample. The total contribution from 
all sources is predicted to amount to $(7.9\pm0.5)\%$, while the fit in 
the Monte Carlo sample measures this to be $(7.8\pm2.2)\%$. 
In the data the same procedure gives the contribution 
from partially reconstructed decays
to be $(8.1\pm1.7)\%$, in excellent agreement. 
In total $8497$ candidates are found, of which 
$3750 \pm 24$ are background events, where the error 
given is the statistical error from the fit. 

\subsection{Flavour Composition and Fragmentation Fits}

The main contributions to the sample of tagged events are 
the same as described in section~\ref{sec-flavsep}. 
The principal method for determining the 
flavour composition is the 
same as was described for the exclusive $\PDstp$ sample.
The goal of this analysis 
is the determination of the absolute rate of 
$\PDstp$ production in charm and bottom decays. Therefore the 
observed number of $\PDstp$ mesons needs to be 
corrected for the reconstruction efficiency
which can be done reliably only in the 3-prong sample.
In addition knowing the efficiencies allows to 
constrain the shape of the fragmentation function 
to a particular shape. 
Previous studies~\cite{bib-OPALD*,bib-OPALcharm} 
have shown that the efficiency corrected fragmentation 
function in charm decays can be described well by 
the function of Peterson et al.~\cite{bib-PETERSON}. 
\begin{figure}[t]
\begin{center}
  \epsfig{file=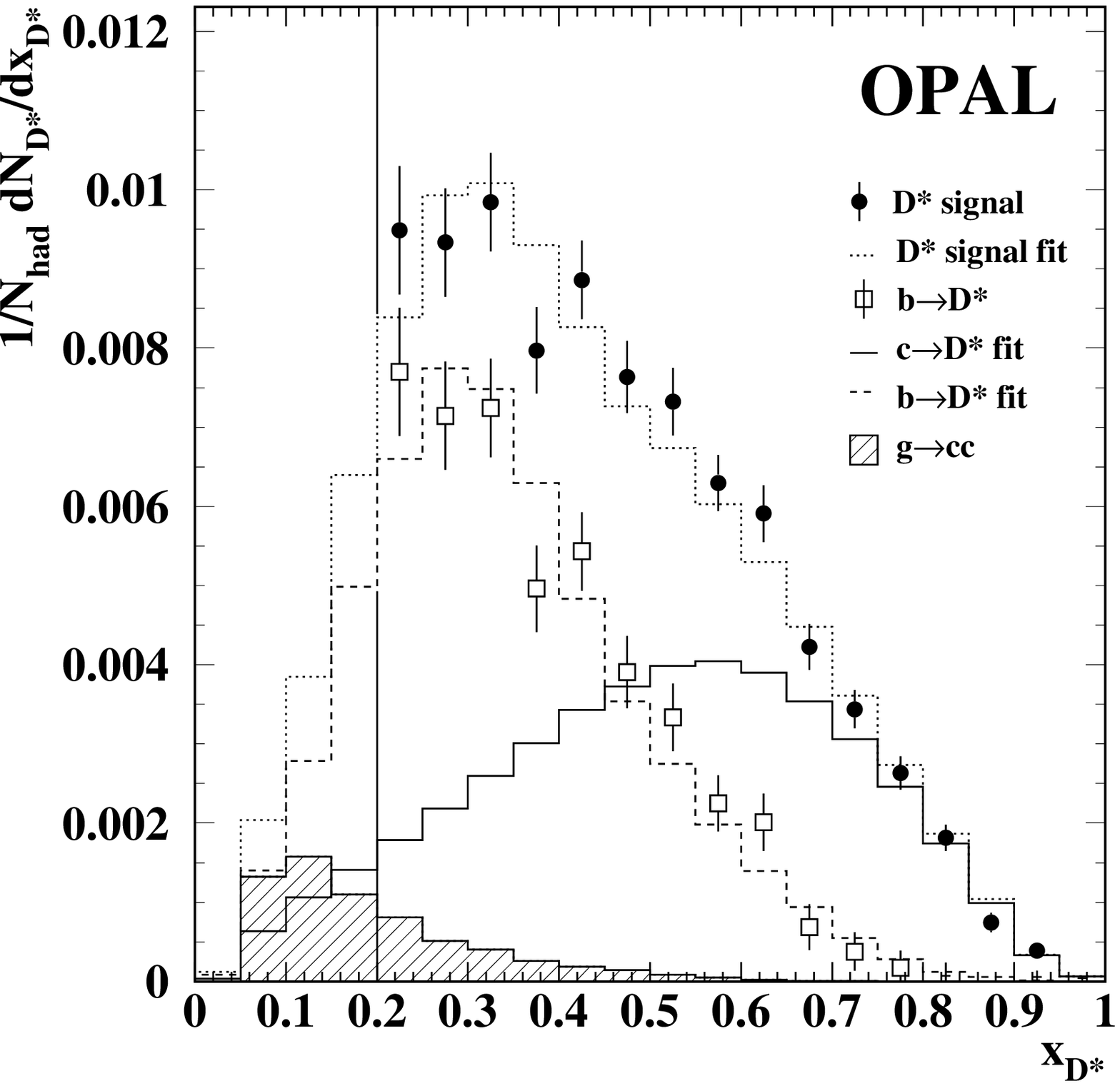,height=12cm}
  \smcap{\label{fig-dndx} Efficiency corrected yield 
of $\PDstp$ mesons as a function of the scaled 
energy $x_{\PDstp}$, 
\protect $1/N_{\rm had} {\mathrm d}N_{\PDstp}/{\mathrm d}x_{\PDstp}$, for all 
candidates (filled points with error bars) reconstructed in the 
decay $\PDstp\to\PDz\Pgp, \PDz\to\PK\Pgp$, and the charm component 
after flavour separation (solid line). The open points 
are the bottom component, and the dashed line 
represents the result of the fit.
Also shown is the predicted 
contribution from gluon splitting events (hatched area). The reconstruction 
is only done for $x_{\PDstp}>0.2$, as indicated by the solid vertical line.}
\end{center}
\end{figure}

The flavour separation is done in 
the $x_{\PDstp}$ range
$0.2<x_{\PDstp}<1.0$, subdivided into 16 bins of $x_{\PDstp}$. 
The charm and the bottom fragmentation functions are constrained 
to the Peterson shape  
by convoluting the analytical fragmentation function with 
the effects of the hadronisation as predicted
by the JETSET 7.4 model. 
A simultaneous fit is then done using the mean scaled 
energy and the information from the flavour separation procedure. 
The normalisations for both bottom and charm decays
and the Peterson parameters $\varepsilon_{\Pc}$ and $\varepsilon_{\Pb}$ 
are allowed to vary in the fit. 

The efficiency of the $\PDstp$ reconstruction is 
calculated in bins of $x_{\PDstp}$
separately for $\PZz\to\bbbar$ and 
for $\PZz\to\ccbar$ events  in the Monte Carlo simulation. 
The efficiency is essentially constant as 
a function of $x_{\PDstp}$, with a small step at $x_{\PDstp}=0.5$ 
due to the change in the $\dEdx$ and the $\cos \theta^*$ cuts. 
Typically it is $(25.0\pm0.6)\%$ for $x_{\PDstp}<0.5$, 
and $(30.0\pm0.5)\%$ for $x_{\PDstp}>0.5$, with the 
bottom and charm efficiencies being very similar. 

Some of the tagged $\PDstp$ events 
are expected to come from gluon 
splitting. The OPAL measurement for the gluon splitting 
rate is used~\cite{bib-OPALD*,bib-OPALgcc}, 
and the shape is taken from the Monte Carlo prediction 
as was done in~\cite{bib-OPALD*}. 
The result of the separation is 
shown in figure~\ref{fig-dndx}, where the total 
efficiency-corrected yield, the charm component,
the bottom component and the part from gluon splitting are shown.

Integrating the fitted fragmentation functions over the 
full $x_{\PDstp}$ range  
the product branching ratio in charm events is found to be
\begin{eqnarray}
  \Gcc/\Ghad \cdot {\fctoD} \;
  \BR(\PDstp\to\PDz\Pgp)\;\BR(\PDz\to\PK\Pgp)  
  &=& ( \nZctoD \pm \estatZctoD ) \times 10^{-3} \ , \nonumber
\end{eqnarray}
where only the statistical error has been quoted. The $\chi^2$
per degree of freedom is $1.28$.

Though not directly needed in this analysis the same 
fit returns information about the production of 
$\PDstp$ mesons in $\PZz\to\bbbar$ events, the 
total $\PDstp$ multiplicity, and the hardness of 
the $\PDstp$ fragmentation function in charm decays. 
In bottom events the product branching ratio is determined to be 
\begin{eqnarray}
  \Gbb/\Ghad \cdot {\fbtoD} \;
  \BR(\PDstp\to\PDz\Pgp)\;\BR(\PDz\to\PK\Pgp)  
  &=& ( \nZbtoD \pm \estatZbtoD ) \times 10^{-3} \ . \nonumber
\end{eqnarray}
The errors quoted are only statistical. The correlation 
between the total production rate in $\PZz\to\ccbar$ and 
in $\PZz\to\bbbar$ events is found to be $-23\%$.
Adding the predicted gluon component, and correcting 
for the branching ratios 
$\BR(\PDstp\to\PDz\Pgp)=0.683 \pm 0.014$ and 
$\BR(\PDz\to\PK\Pgp)=0.0383\pm0.0012$~\cite{bib-PDG},
the total multiplicity 
of $\PDstp$ mesons in hadronic $\PZz$ decays is found to be 
\begin{equation}
  \bar n_{\PZz\to\PDstp X} = 
 \nZtoD \pm \estatnZtoD \nonumber\ .
\end{equation} 
In addition 
the shape of the fragmentation function allows the 
determination of the average $x_{\PDstp}$ between $0$ and $1$ 
in charm decays to be measured as
\begin{equation}
  \langle x_{\PDstp} \rangle_{\Pc} = \xc \pm \estatxc \ .
\nonumber
\end{equation}
Again only the statistical error is quoted. 
Note that this measurement does not include the effects 
from $\PDst$ mesons produced in gluon splitting. 

\subsection{Systematic Errors of the Measurement of 
\boldmath $\PDstp$ Production}
\label{sec-dstar-syserr}

A number of systematic errors are investigated in connection 
with the $\PDstp$ rate measurements in charm and in bottom 
events. Note that the errors are totally anticorrelated, 
and only the ones for the rate in $\PZz\to\ccbar$ events are 
given. Errors of the determination 
of the mean scaled energy of $\PDstp$ mesons are discussed separately 
at the end of this section. 
The first group of errors are due to detector effects, 
resolutions and Monte Carlo modelling:
\begin{itemize}
\item Track quality cuts: The effects of the track quality cuts 
are investigated by comparing the efficiency for each 
selection cut in data and Monte Carlo. 
An error of $\pm 0.6\%$ has 
been found to be sufficient to cover observed differences.
\item Fraction of silicon hits: 
The resolution of secondary vertices depends on 
the fraction of tracks which use measurements from the 
silicon micro-vertex detector. 
The fraction in Monte Carlo events has been re-weighted to 
the one measured in data. An error of $\pm 0.4\%$ is 
derived from the statistical precision of this procedure. 
\item $\dEdx$ modelling: The calibration of the specific energy 
loss, $\dEdx$, has been compared in data and Monte Carlo in 
samples of identified particles. Samples of kaons and 
pions are selected in decays of $\phi\to\PK\PaK$ and 
$\mathrm K^0\to\Pgp\Pagp$ mesons
without using $\dEdx$ requirements, 
and the calibration of $\dEdx$ is measured in the data. 
In addition $\PDstp$ mesons are reconstructed without 
$\dEdx$ requirements, and the results are compared with 
those quoted in table~\ref{tab-NDST}.
For the applied cut of $2\%$ on the kaon weight an 
error of $1.1\%$ of the total rate was found in those candidates 
where a cut was applied. This error contains a 
contribution from the measurement of $\dEdx$ itself, mainly 
due to the calibration, and from the requirement of 
at least $20$ hits for the $\dEdx$ measurement. 
Since a $\dEdx$ cut is only used for 
$x_{\PDstp}<0.5$ this translates into a reduced error on the total rate in 
charm events of $0.8\%$, and in bottom events of $1.0\%$.
%
%
%
\item Mass resolution: The invariant mass resolutions in data and 
Monte Carlo for the decay selected have been compared. The 
$M_{0}$ resolution in 
the Monte Carlo is $27.5~\MeV$, the one in data $27.9~\MeV$. 
Depending on the Monte Carlo sample used variations in the 
mass resolution from sample to sample of up to a few $\MeV$
are observed. These differences correspond to 
changes in the momentum resolution of the detector of 
approximately $10\%$, and translate into a systematic 
error on the efficiency of $1.0\%$. 
\item Rate of partially reconstructed $\PDstp$: 
The total contribution from partially reconstructed $\PDstp$ mesons 
is measured in the fit
to be $(8.1\pm1.7)\%$ in the selected sample. 
Of these $(4.6\pm0.8)\%$ are reconstructed as coming 
from the satellite decay mode. The Monte Carlo 
simulation predicts the fraction of partially 
reconstructed $\PDstp$ decays to be 
$(7.9\pm0.5)\%$, of which \mbox{$(4.8 \pm 0.4)\%$} are 
from the satellite decay mode. The rest, $(3.3\pm0.3)\%$, 
are from a number of other decays, as discussed 
in section~\ref{sec-excl-bgd}. In general very 
good agreement is observed between the predicted and 
the measured fractions, both within the Monte Carlo 
simulation, and between data and Monte Carlo. 
The uncertainty of the method is estimated from the 
statistical precision of the fit, and from 
the error of the individual branching ratios 
contributing to the partially 
reconstructed signal in the Monte Carlo. In 
addition a contribution
of $0.5\%$ is included to account for the finite 
Monte Carlo statistics available for this study. 
In total the relative error of the rate 
of partially reconstructed $\PDstp$ mesons contributing
to the signal is estimated to be $22\%$ of the 
rate of partially reconstructed mesons, which contributes
an error of $1.5\%$ to the total rate measurement.
\item Background subtraction: The combinatorial background 
in the sample is subtracted based on estimators derived from 
data. The quality 
of the procedure has been studied in the Monte Carlo simulation. 
Within the available statistics no significant deviations are found. 
The total difference is less than $1\%$, and an error of 
$1\%$ is assigned to this source. 
\item $\mathrm{g}\to\ccbar$: The expected contribution from gluon splitting 
has been subtracted from the sample, for $x_{\PDstp}>0.2$. 
The mean value used is the 
one measured in \cite{bib-OPALD*,bib-OPALgcc} of \mbox{$0.0238\pm0.0048$}.
The contribution from this process has been varied within this error, 
which results in an error on the total rate of $\pm 1\%$.
Monte Carlo studies indicate that the shape of the $\mathrm{g}\to\ccbar$
component is not very dependent on the particular Monte Carlo model used. 
Comparing JETSET and the Ariadne Monte Carlo model \cite{bib-ARIADNE} an error 
of $\pm 1\%$ is assigned to this source, resulting in a total error 
from gluon splitting of $\pm 1.4\%$. 
\item Heavy flavour fragmentation: 
The total rate of $\PDstp$ 
mesons produced in $\PZz\to\ccbar$ and in $\PZz\to\bbbar$ 
decays has been obtained by 
integrating the fitted fragmentation functions from $x_{\PDstp}=0$ to $1$. 
This procedure is subject to a number of systematic uncertainties:
\begin{itemize}
  \item fragmentation function: 
The Peterson fragmentation function has been used for the results 
quoted. The different fragmentation 
models of Collins and Spiller~\cite{bib-Collins} 
and Kartvelishvili~\cite{bib-Kart} 
have been used to estimate its influence.
The largest difference in the fitted rate 
has been used as the systematic error from this source. 
The error found is $1\%$ of the rate in charm events, 
and $4\%$ in bottom events. 

As a cross-check the fits have been repeated using the 
QCD inspired fragmentation function from Nason \etal \cite{bib-NASON}. 
The results are compatible with the ones obtained 
using the fragmentation models. 
The error on the fitted parameters however also reflect 
the limited statistics available for the initial 
tune of the model using low energy data. 
It is therefore used as a cross-check 
rather than to give an additional error. 

\item b decay modelling: A comparatively large fraction of $\Pb\to\PDstp$ 
decays are not observed since only events with $x>0.2$ are 
considered in this analysis. In addition to 
the error from different fragmentation models discussed 
above, an error introduced by this 
extrapolation on the measured b-rate has been studied as 
in~\cite{bib-OPALcharm} by considering the different 
types of decays contributing to the spectrum, 
and investigating the differences in the extrapolation 
introduced by each of the components. 
The differences found 
amount to an error of $2\%$ on the $\Pb\to\PDst$ rate measurement. 
%
\item excited D meson production:
The shape of the fragmentation functions is 
influenced by the presence of D excited states in the decay chains. 
About $32\%$ of all $\PDstp$ mesons have been measured 
as originating in decays of excited charm mesons. 
The contribution from these has been varied by $\pm 18\%$ around 
the mean value of $32\%$, as was done in~\cite{bib-OPALcharm}. 
The resulting error is 
$0.2\%$ on the rate. 
\end{itemize}
The total error from all fragmentation modelling issues is 
$3\%$ for charm, $5\%$ for bottom. 
\end{itemize}
In addition the errors discussed in the section on the 
flavour separation apply to the rate measurements as well.
A complete 
breakdown of errors for the measurements is given in 
table~\ref{tab-syserrDrate}.
\begin{table}[p]
\begin{center}
  \begin{tabular}{|l|r@{\ \ }|r@{\ \ }|r@{\ \ }|r@{\ \ \ }|}
  \hline
  & charm    & bottom &
          \protect$\overline n_{\PZz\to\PDstp}$& 
\multicolumn{1}{c|}{$\langle x_{\PDstp} \rangle_{\Pc}$}\\
  \hline\hline
\multicolumn{1}{|c|}{    error source}  & \multicolumn{2}{c|}{$\times 10^{-3}$}&
\multicolumn{2}{c|}{}\\
\hline \hline
\multicolumn{5}{|l|}{ {\it detector resolution}}\\
\hline
    track quality cuts     &    0.006  & 0.008 & 0.0008 & \\
    fraction of silicon hits&    0.003  & 0.004 & 0.0004 & 
\raisebox{0mm}[0mm][0mm]{{$\left\}\begin{array}{r}\\ \\ \\ \end{array}\right.$}}\hfill 0.001 \\     
    $\dEdx$ modelling      &    0.008  & 0.013 & 0.0012 & \\
\hline
    total detector resolution & 0.011 & 0.016 &  0.0015& 0.001 \\

\hline\hline
\multicolumn{5}{|l|}{{\it $\PDstp$ reconstruction}}\\
\hline
    mass resolution        &    0.010  & 0.013 & 0.0013 & \\
    subtracting partially reconstructed decays
                           &    0.016  & 0.020 & 0.0019 & 
\raisebox{0mm}[0mm][0mm]{{$\left\}\begin{array}{r}\\ \\ \\ \end{array}\right.$}}
\hfill 0.002 \\     
    $\PDstp$ background subtraction &    0.010  & 0.013 & 0.0013 & \\
    g$\to\ccbar$           & $-$0.016  & $-$0.019 & $-$0.0018 & 0.004\\
\hline
total $\PDstp$ reconstruction & 0.026 & 0.030 & 0.0032   &  0.005 \\
\hline\hline
\multicolumn{5}{|l|}{{\it flavour separation}}\\
\hline
    $f_{\Pb}$ statistical error   &    0.010  & 0.013 & &\\
    background modelling   &    0.017  & 0.023 & &\\
    hemisphere correlation &    0.004  & 0.005 & &\\
    charm modelling        &    0.010  & 0.013 & & \\
    bottom multiplicity    & $-$0.005  & $+$0.007& & 
\raisebox{0mm}[0mm][0mm]{{$\left\}\begin{array}{r}\\ \\ \\ \\ \\ \\ \\ \\ \\ \end{array}\right.$}}
\hfill 0.002 \\     
    charm multiplicity     & $+$0.006  & $-$0.008 & & \\
    bottom lifetime        & $-$0.007  & $+$0.009 & & \\
    charm lifetime         & $+$0.003  & $-$0.004 & & \\
    B mixing               &    0.003  & 0.004 & & \\
    fragmentation model    &    0.012  & 0.053 & 0.0041 &0.007 \\
    b decay model          &           & 0.027 & 0.0021 & \\
    excited D meson production&   0.002  & 0.003 & 0.0003 &0.002\\

\hline
total flavour separation   & 0.030 & 0.070 & 0.0047 &  0.008 \\
\hline\hline
    total       & \protect$\esysZctoD$ & \protect$\esysZbtoD$ & 
                         \protect$\esysnZtoD$&  \protect$\esysxc$\\
\hline
\end{tabular}
\smcap{\label{tab-syserrDrate} List of systematic errors relevant 
in the determination of the rate of \protect$\PDstp$ mesons produced in 
\protect$\PZz\to\ccbar$ and in \protect$\PZz\to\bbbar$ events, the 
total multiplicity of \protect$\PDstp$ mesons in \protect$\PZz$ 
decays, and the mean scaled energy. For the error of the 
mean scaled energy only errors which are larger 
than $0.001\times 10^{-3}$ are listed separately, otherwise the total 
contribution from a class of errors is given.  
A sign in front of an error indicates the direction of change   
under a positive change of the variable. The last three errors 
from the flavour separation include the contribution from 
the same sources through the $\PDstp$ reconstruction. 
Note that all flavour 
separation errors are anti-correlated between the charm and the 
bottom result.}
\end{center}
\end{table}

Only a few of the errors listed for the rate measurement 
have a significant effect on the determination of 
the mean scaled energy of $\PDstp$ mesons in charm decays. 
The most important error comes from the extrapolation 
of the fragmentation function into the unmeasured region 
below $x_{\PDstp}=0.2$. Making the same comparisons between 
different fragmentation models as described 
above a modelling error of $\pm 0.007$ for the mean scaled energy 
has been determined. The uncertainty in the modelling of 
gluon splitting results in a further error 
of $0.004$.
Effects related to the flavour separation 
have much less of an effect on the mean scaled energy. 
Taken together all other errors except the above-mentioned 
modelling issues contribute another $0.004$ to the error. 
The reconstruction of $\PDstp$ mesons, in particular the 
background subtraction, contribute an additional overall error 
of $0.002$. 

The final results, including all systematic errors, for the 
hadronisation fractions of charm and bottom quarks into 
$\PDstp$ mesons are found to be 
$$ \Gcc/\Ghad \cdot {\fctoD}\;\BR(\PDstp\to\PDz\Pgp)\;
\BR(\PDz\to\PK\Pgp) = 
  (\nZctoD \pm \estatZctoD \pm \esysZctoD ) \times 10^{-3}\ ,$$ and 
$$ \Gbb/\Ghad \cdot {\fbtoD}\;\BR(\PDstp\to\PDz\Pgp)\;
\BR(\PDz\to\PK\Pgp) = 
  (\nZbtoD \pm \estatZbtoD \pm \esysZbtoD ) \times 10^{-3}\ , $$
where the first error is statistical, the second systematic.
The total multiplicity 
of $\PDstp$ mesons in hadronic $\PZz$ decays is found to be 
$$
  \bar n_{\PZz\to\PDstp X} = 
 \nZtoD \pm \estatnZtoD \pm \esysnZtoD \pm \esysnZtoDBR \nonumber\ .
$$
The last error quoted is due to external branching ratios.
From the shape of the fragmentation function the average mean scaled 
energy $x_{\PDstp}$ of $\PDstp$ mesons in charm decays is determined to be 
$$
  \langle x_{\PDstp}\rangle_{\Pc} = \xc \pm \estatxc \pm \esysxc \ .
\nonumber
$$
The errors quoted are statistical and systematic, respectively. 
The results quoted is for the primary production of 
$\PDstp$ mesons in $\PZz\to\ccbar$ events, and 
does not contain contributions from gluon splitting events. 
The results are in agreement with other measurements at LEP 
and with the previous OPAL determination 
\cite{bib-OPALD*,bib-ALEPHD*,bib-DELPHID*}.

\section{\boldmath 
Measurement of ${\fctoD}$, ${\fbtoD}$ \\ and of 
$\Gamma_{\ccbar}/\Gamma_{\mathrm had}$}
\label{sec-fctoD}
The hadronisation fraction ${\fctoD}$
is measured by the simultaneous detection of charm quarks in 
two jets of the event. In one jet, a charm meson candidate is 
identified by using the exclusive $\PDstp$  meson tag discussed in 
section~\ref{sec-dstar}. In the other jet  
the inclusive tag presented in section~\ref{sec-slowpion} is used. 
Background is suppressed by requiring that both $\PDstp$ tags 
are of opposite charge where the charge of the $\PDstp$ is given through
the charge of the slow pion candidate from the $\PDstp\to\PDz\Pgp$ decay.
By determining both the number of exclusively reconstructed 
D mesons and of simultaneously tagged jets, the hadronisation 
fraction can be calculated. 

The hadronisation fraction $\fbtoD$ is measured 
in an analogous fashion by the simultaneous detection of a 
lepton in one jet and an inclusively reconstructed 
$\PDstp$ meson in another jet of the event. 
The exclusive charm tag 
is replaced by a lepton tag, optimised 
for the selection of bottom events. The inclusive tag based 
on the slow pion is used in the opposite jet.

Combining the hadronisation fraction $\fctoD$ with the 
measurement of the production of $\PDstp$ mesons 
in $\PZz\to\ccbar$ events the relative charm 
partial width $\Gamma_{\ccbar}/\Gamma_{\mathrm had}$ is 
calculated, as discussed at the end of this section. 

\subsection{The Tagged Samples}
\label{sec-tagged}
Two samples are used in this analysis, one tagged 
by the presence of an exclusively reconstructed $\PDstp$ meson,
the other tagged by a lepton. Both are selected as 
described in section~\ref{sec-ctag}. After all cuts 
a sample of $27\thinspace 005$ $\PDstp$ candidates has been found, of which 
$11\thinspace 366\pm107$ are estimated background events. 
Of the selected $\PDstp$ events a fraction of $0.774 \pm 0.023$ 
are reconstructed as coming from $\PZz\to\ccbar$ 
events. 
The number of tagged 
electron and muon candidates
is determined to be $43\thinspace 579$, 
of which $4445\pm64$ do not originate in bottom decays. 


Slow pions are sought in the secondary jet
in the samples of events tagged by either a $\PDstp$ meson
or a lepton. The reconstruction of 
slow pion candidates proceeds as described in section~\ref{sec-slowpion}. 
Event samples are prepared for the correct charge combination 
$\PDstp$-$\pi^-$ and for any charge combination $\ell$-$\pi$, respectively.
Background in the slow pion sample is estimated from 
a $\PDstm$-$\pi^-$ sample, prepared using a selection in a sideband 
of $\Delta M>0.18$.
A clear enhancement is visible at low values of $p_t^2$ 
as shown in figure~\ref{fig-slowpi}(a) for the $\PDstp$ tag, 
and in figure~\ref{fig-slowpi}(b) for the lepton tag. 
A much smaller enhancement is visible in the sideband selected double 
tag, shown in figure~\ref{fig-slowpi}(c).
The number of double tagged candidates, counted below 
$p_t^2<0.01\; \GeV^2$, is $2146$ $\PDstp$-$\Pgp$ candidates, and 
$5502$ $\ell$-$\pi$ candidates. 

The efficiency to reconstruct a slow pion in the 
presence of a fully reconstructed $\PDstp$ meson or a lepton in the 
other jet is determined in Monte Carlo simulated events. 
The slow pion is sought
in the secondary jet of the event, and its reconstruction efficiency 
is calculated. 
This procedure is done individually in each of the 
five exclusive $\PDstp$ modes, and for the lepton tagged sample. 
The final efficiency is 
calculated by reweighting the Monte Carlo efficiencies to reflect the 
mixture of tagged events in the data.
After all cuts the efficiencies are found to be 
\begin{equation}
    \epsilon_{\PDst\pi}^{\Pc} = 0.384 \pm 0.004 \ \ {\mathrm and}\ \ 
    \epsilon_{\PDst\pi}^{\Pb} = 0.189 \pm 0.006 \ , 
\end{equation}
for $\Pc\to\PDstp$ and $\Pb\to\PDstp$ events, respectively. In 
lepton tagged events the efficiencies are 
\begin{equation}
   \epsilon_{\ell\pi}^{\Pc} = 0.414 \pm 0.029 \ \  {\mathrm and} \ \ 
   \epsilon_{\ell\pi}^{\Pb} = 0.193 \pm 0.005 \ .
\end{equation}
The errors quoted are purely statistical.

\subsection{Composition of  the Double Tagged Sample}
\label{slowpiback.sec}
A number of different classes of events contribute to the 
sample of double tagged events. At very low $p_t^2$ 
a significant fraction of the candidates are due to slow pions 
from the decay $\PDstp\to\PDz\Pgp$, both in charm and in bottom decays. 
The signal in bottom decays, while similar to the one in charm, 
has a broader distribution in $p_t^2$. 
Background in the sample comes from a number of different processes.
The dominant source is random tracks that pass the 
applied cuts. This combinatorial background 
falls significantly more slowly with increasing $p_t^2$ than does the signal, 
and does not exhibit the characteristic enhancement at very low $p_t^2$.


A small but important background is slow pions from 
fake double tag candidates. 
A double tagged event is denoted a ``fake double tag'', 
if the slow pion candidate found in the 
inclusive tag is correctly identified, but the fully 
reconstructed $\PDstp$ meson or the lepton in the other jet 
is wrongly identified. 
Such events contribute 
to the peak in the $p_t^2$ spectrum, and need to be subtracted from 
the sample. 

In the following each of the different parts of the candidate distribution 
will be briefly discussed. In the last part of this section 
the method used to count the number of double tags from 
charm decays is presented. 

\subsubsection{Combinatorial Background and Fake Estimation}
\label{sec-slowpi-bgd}
The combinatorial background is the dominant background source. 
Its shape is estimated using events with the wrong charge 
correlation between the $\PDstp$ and the slow pion in the 
opposite hemisphere, which 
in addition are reconstructed in a sideband of $\Delta M$ 
on the exclusive side,  
between $0.18~(0.19)~\GeV < \Delta M < 0.20~(0.22)~\GeV$ (numbers 
in brackets are for the semileptonic channels). 
Searching for slow pions in the secondary jet relative to the
exclusive candidates tagged in the sidebands, a 
signal for charm production is observed at low $p_t^2$
(figure~\ref{fig-slowpi}(c)). 
This signal has two contributions, 
one from fake double tags,
another small one from incompletely reconstructed $\PDstp$ meson decays 
in the D-jet. In both cases a true slow pion is found in the secondary jet. 
The total fraction of fake double tags in the background 
is measured in the sidebands, and is subtracted from the 
total number of double tags, as described in~\ref{sec-ndouble}.

\subsubsection{Contribution from Bottom Events}
\label{sec-Bmix}
The shape of the $p_t^2$ signal in bottom events is
determined in data from the lepton-slow pion double tagged 
sample, which is about $90\%$ pure in bottom decays. 
The fraction of events in the $\PDstp$-$\Pagp$ double tagged sample 
originating from bottom decays is determined from the 
known fraction of b-events in the $\PDstp$ sample, and 
the efficiency to tag a slow pion in a b-decay in the 
secondary jet. 

The situation is slightly complicated by 
mixing in the neutral B system.
%
If mixing has occurred in either hemisphere, the charge correlation 
between the primary quark and the corresponding  $\PDstp$ mesons 
is changed, and the correlation between the charge of the 
slow pion track and the fully reconstructed $\PDstp$ candidate 
is opposite
to the unmixed case. This fraction of events 
migrates out of the signal sample in this analysis.
The total probability in bottom events that mixing has destroyed the charge 
correlation is given by 
\begin{eqnarray}
\chi_{\rm eff}^{\PDstp} &=& \chi_{\PDstp}
(1-\chi_{{\Pgp}_{\rm slow}}) +
                 \chi_{{\Pgp}_{\rm slow}}(1-\chi_{\PDstp})
\end{eqnarray}
where $\chi_{\Pgp_{\mathrm {slow}}}, \chi_{\PDstp}$ are the 
effective mixing parameters applicable to the slow 
pion and the $\PDstp$ sample, respectively. The two 
mixing parameters are equal, since a $\PDstp$ tag is used in both jets. 
The majority of $\PDstp$ mesons in 
$\PZz\to\bbbar$ events 
originate from decays of the $\PBz$ meson. This fraction is 
estimated using
semileptonic B decays to be  $r_{\Pd} = 0.790 ^{+0.13}_{-0.12}$ 
\cite{bib-mix1}. A small percentage of $\PDstp$ mesons are also
expected from $\PB_{\mathrm s}$ mesons, which also mix.
The number of $\PDstp$ from  $\PB_{\mathrm s}$ mesons was estimated to be
$r_{\Ps} = 0.033 \pm 0.015$. 
The average mixing in the neutral B system is determined from the 
world average value for 
the mixing parameter, $\chi_{\Pd} = 0.175\pm 0.016$ \cite{bib-PDG}.
For the mixing parameter of the $\PB_{\mathrm s}$ meson 
the current world average limit of 
$\chi_{\Ps}>0.49 \; {\mathrm {at}} \; 95 \%$ confidence level~\cite{bib-PDG}
has been used.
In this analysis $\chi_{\Ps}$ 
is varied between 0.49 and 0.50. 
The effective mixing seen by the $\PDstp$ mesons is given by 
\begin{equation}
    \chi_{\PDstp} = r_{\rm d} \; \chi_{\rm d} + r_{\rm s} \; \chi_{\rm s}.
\end{equation}
In addition $\PDstp$ mesons with the wrong sign can be produced 
in bottom decays, when a $\Pac$ quark is produced in the decay of 
the W. This can be expressed in terms of a mixing-like parameter 
$\zeta_{\rm D}$. As in \cite{bib-OPALD*} a value of 
$\zeta_{\rm D}=0.025 \pm 0.025$ is used. 
The effective mixing parameter for the $\PDstp$-$\Pagp$ 
double tag sample is estimated to be
\begin{equation}
    \chi^{\PDstp}_{\mathrm eff} = 0.289 \pm 0.050. \nonumber
\end{equation}   
This number is in agreement with a direct measurement of the effective 
$\PDstp$ mixing in \cite{bib-OPALD*AFB}.


\subsection{Determination of the Number of Double Tags}
\label{sec-ndouble}
The number of double tagged events in the sample 
is estimated from the three $p_t^2$ distributions by 
a simultaneous fit. The right sign sample is 
fitted as a superposition of true signal from charm and 
bottom decays, a contribution from fake double tagged 
events, and background. 
Each of the individual contributions is described by an exponential 
function 
\begin{equation}
F(p_t^2)_{j=\rm bgd,b,c} = a_j \; \exp ( b_j p_t^2 ) \ .
\end{equation}

The parameters of the signal 
originating from bottom decays is determined in the 
$\ell$-$\pi$ double tagged sample. 
The fake distribution is measured 
in a double tagged sideband sample, where the exclusive candidates 
are selected in a sideband of $\Delta M$. 
Combinatorial background is fitted for in the different 
distributions. 

All three distributions are simultaneously fitted using the 
following parametrisations.
The shape of the lepton tagged spectrum is parametrised by
\begin{equation}
F(p_t^2)_{\ell\pi} = F(p_t^2)_{\rm bgd} + 
f_{\Pb}^{\ell} \;{\fbtoD} \; F(p_t^2)_{\Pb}\; \epsilon_{\ell\pi}^{\Pb} +
f_{\Pc}^{\ell} \;{\fctoD} \; F(p_t^2)_{\Pc}\; \epsilon_{\ell\pi}^{\Pc} \ ,
\label{eq-pt2l}
\end{equation}
where the purities $f_{\Pc}^{\ell}$ and $f_{\Pb}^{\ell}$ have been 
given in section~\ref{sec-lepttag}, and 
$\epsilon_{\ell\pi}^{\Pb}, \epsilon_{\ell\pi}^{\Pc}$ are the 
efficiencies to find a slow pion in the secondary jet in the 
presence of a lepton in the other jet, in $\bbbar$ and $\ccbar$ events 
respectively, as quoted in section~\ref{sec-tagged}. 
Since both relative signs between leptons and 
pions are used eq.~\ref{eq-pt2l} does not depend on 
the mixing in the lepton tagged sample.
The right sign signal distribution is described by
\begin{eqnarray}
\label{eq-pt2sig}
F(p_t^2)_{\PDst\pi} =&& F(p_t^2)_{\rm bgd} + 
        F(p_t^2)_{\mathrm fake} \nonumber\\
    & + &(1- \chi_{\rm eff}^{\PDstp}) \; (1-f_{\Pc}^{\PDstp})\;
      {\fbtoD } \;  F(p_t^2)_{\Pb}\;\epsilon_{\PDst\pi}^{\Pb} \\
    & + &f_{\Pc}^{\PDstp} \; 
      {\fctoD}  \; F(p_t^2)_{\Pc}\; \epsilon_{\PDst\pi}^{\Pc} \nonumber\ .
\end{eqnarray}
The charm fraction $f_{\Pc}^{\PDstp}$ fulfils the condition 
$f_{\Pc}^{\PDstp} = (1 - f_{\Pb}^{\PDstp})$, 
and $\epsilon_{\PDst,\pi}^{\Pc}$, $\epsilon_{\PDst,\pi}^{\Pb}$ are the 
efficiencies to find a slow pion in the 
presence of a $\PDstp$ meson in the other jet in $\ccbar$ and $\bbbar$ 
events respectively. The mixing probability 
$\chi_{\rm eff}^{\PDstp}$ has been determined in section~\ref{sec-Bmix}. 

The contribution from fake double tags in the $\PDstp-\Pagp$ 
double tagged sample 
is measured in the sideband tagged sample, as described above.
The $p_t^2$ spectrum in the sideband tagged sample 
is parametrised by 
\begin{equation}
F(p_t^2)_{\rm side} = \alpha\; ( F(p_t^2)_{\rm bgd} + 
      \; F(p_t^2)_{\mathrm fake} )\ ,
\label{eq-pt2WQ}
\end{equation}
where $F(p_t^2)_{\mathrm fake}$ contains contributions 
from fakes in the double tagged sample and from partially reconstructed 
$\PDstp$ mesons. The contribution from fake double tags 
is assumed to have the flavour composition as the real 
signal, as given by the two last lines of eq.~\ref{eq-pt2sig},
and the same functions are used for $F(p_t^2)_{\mathrm fake}$
as for signal events. 
%
The absolute 
contribution from the fakes is obtained by 
rescaling the fitted fake contribution by the ratio $\alpha$
of the number of background candidates in the 
sideband sample to that in the signal sample. 
%

\begin{figure}[p]
  \begin{center}
  \epsfig{file=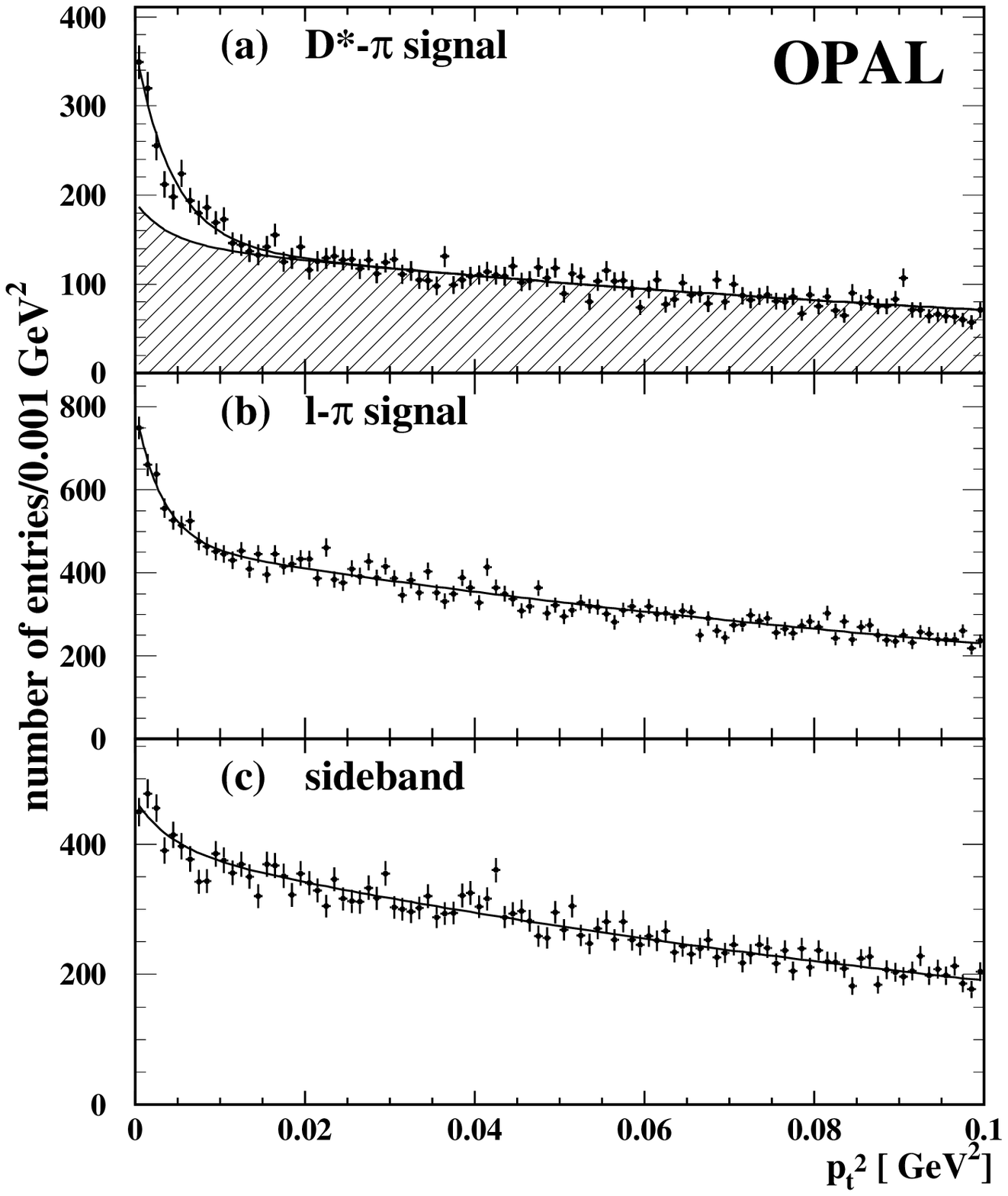,height=20cm}
\smcap{\label{fig-slowpi}
Spectrum of the squared transverse momentum of double tagged 
candidates after applying all cuts of the 
(a) signal candidates, reconstructed with opposite charges in both jets.
The non-charm component is indicated by the hatched area;
(b) lepton-tagged candidates. The points are the data and the lines 
the results of the fit;
(c) background candidates, reconstructed in \protect$\Delta M$ sidebands.
}
  \end{center}
\end{figure}

Equations \ref{eq-pt2l}, 20 and \ref{eq-pt2WQ} are 
fitted simultaneously. Free parameters in the fit are 
the hadronisation fractions $\fctoD$ and 
$\fbtoD$, the normalisation of the 
fake rate, the shape parameters $a$ and $b$ of bottom and charm slow pion 
signals, and the background parameters. 
The different spectra are 
illustrated in figures~\ref{fig-slowpi}(a) to (c), with the results 
of the fit superimposed in each case. 
The number of double tagged events in the $\PDstp$-$\Pagp$ double tagged 
sample in $\PZz\to\ccbar$ events, $N_{\PDstp,\Pagp}^{\Pc}$,
and in the $\ell$-$\pi$ tagged sample, $N_{\ell,\pi}^{\Pb}$,
are determined by integrating the fitted signal functions 
in the double tagged samples between $p_t^2=0\;\GeV^2$
and $p_t^2=0.01\; \GeV^2$. They are found 
to be $\NDP^{\Pc} = 702 \pm 44$, over a background of 
$1444 \pm 18$ events, and 
$\NLP^{\Pb}=\Ntol \pm \eNtol$ over a background of $4568 \pm 45$ events, 
respectively.
Correcting for the branching ratio 
$\BR(\PDstp\to\PDz\Pgp)=0.683\pm0.014$~\cite{bib-PDG}
the hadronisation fractions are found to be 
\begin{eqnarray}
{\fctoD} &=& \nctoD \pm \estatctoD \nonumber \\
{\fbtoD} &=& \nbtoD \pm \estatbtoD \nonumber \ ,
\end{eqnarray}
where the error quoted is only statistical. 

Combining the measurement of the hadronisation fraction $\fctoD$
with the result for the total rate of $\PDstp$ mesons 
in $\PZz\to\ccbar$ events the charm partial width relative 
to the total hadronic width of the $\PZz$ is found to be 
$$\Gcc/\Ghad = \nRc \pm \estatRc \ , $$
where the error quoted is purely statistical. 

\subsection{Systematic Errors}
\label{sub-syserr}
In this section the different sources of systematic errors
for this part of the analysis are discussed. A full breakdown 
of all errors considered is given in table~\ref{tab-syserror}.
\begin{table}[p]
\begin{center}
\begin{tabular}{|l|c|c|c|}  
\hline
Source& ${\fbtoD}$&${\fctoD}$&$\Gcc/\Ghad$ \\   
\hline\hline
\multicolumn{4}{|l|}{\it detector resolution} \\
\hline
track quality cuts   &   $\ \ \ $0.0000 & $\ \ \ $0.0000 & $\ \ \ $0.0011\\
$\dEdx$ modelling    &   $\ \ \ $0.0017 & $\ \ \ $0.0025 & $\ \ \ $0.0018\\
detector resolution  &   $\ \ \ $0.0010 & $\ \ \ $0.0013 & $\ \ \ $0.0009\\
\hline\hline
\multicolumn{4}{|l|}{\it flavour separation} \\
\hline
$f_{\Pb}$ stat error  &  $\ \ \ $0.0002 & $\ \ \ $0.0022 & $\ \ \ $0.0025\\
background modelling  &  $\ \ \ $0.0003 & $\ \ \ $0.0030 & $\ \ \ $0.0028\\
hemisphere correlation& $<0.0001$& $\ \ \ $0.0007 & $\ \ \ $0.0007\\
charm modelling       &  $\ \ \ $0.0002 & $\ \ \ $0.0018 & $\ \ \ $0.0017\\
bottom multiplicity   & $<0.0001$        & $-$0.0009 & $-$0.0008\\
charm multiplicity    &  $\ \ \ $0.0001 & $+$0.0011 & $+$0.0010\\
bottom lifetime       &  $\ \ \ $0.0001 & $-$0.0012 & $-$0.0011\\
charm lifetime        &  $\ \ \ $0.0002 & $+$0.0011 & $+$0.0008\\
fragmentation modelling& $<0.0001$      & $\ \ \ $0.0011 & $\ \ \ $0.0018\\
\hline
\hline
\multicolumn{4}{|l|}{\it lepton reconstruction}\\
\hline
lepton fragmentation  &  $\ \ \ $0.0030 & $<0.0001$ &           \\
bottom fraction in $\PZz$ decays & 
                         $\ \ \ $0.0017 & $<0.0001$ &           \\
lepton decay model    &  $\ \ \ $0.0050 & $<0.0001$ &           \\
semi-leptonic BR      &  $\ \ \ $0.0010 & $<0.0001$ &           \\
lepton fake rate      &  $\ \ \ $0.0020 & $<0.0001$ &           \\
\hline\hline
\multicolumn{4}{|l|}{\it $\PDstp$ reconstruction}\\
\hline
mass resolution      &                  &                & $\ \ \ $0.0018\\
subtracting partially reconstructed decays & 
                                        &                & $\ \ \ $0.0027\\
$\PDstp$ background subtraction  &       &                & $\ \ \ $0.0018\\
\hline\hline
\multicolumn{4}{|l|}{\it double tag, slow pion reconstruction } \\
\hline
slow pion efficiency  &  $\ \ \ $0.0040 &  $\ \ \ $0.0010 & $\ \ \ $0.0025\\
excited D meson production  & $\ \ \ $0.0040 &$\ \ \ $0.0010 & $\ \ \ $0.0008\\
 fitting procedure    &  $\ \ \ $0.0081&  $\ \ \ $0.0104 & $\ \ \ $0.0085\\
{$\PBz \overline{\PBz}$ mixing}&$\ \ \ $0.0002&$\ \ \ $0.0016&$\ \ \ $0.0011\\
{jet resolution}      &  $\ \ \ $0.0024 &  $\ \ \ $0.0031 & $\ \ \ $0.0025\\
jet-jet correlation   &  $\ \ \ $0.0012 &  $\ \ \  $0.0015 & $\ \ \ $0.0012\\
{${\rm g}\to\ccbar$}  &  $<0.0001$ &  $-$0.0003 & $-$0.0024\\
\hline\hline
 total error         &  $\ \ \ $ 0.0121 &  $\ \ \ $0.0133 & $\ \ \ $0.0123\\ 
\hline
\hline
\multicolumn{4}{|l|}{\it external inputs}\\
\hline
$\BR(\PDstp\to\PDz\Pgp)$ & $\ \ \ $ 0.0034 & $\ \ \ $ 0.0044 & \\
$\BR(\PDz\to\PK\Pgp)$    &            & & $\ \ \ $0.0063\\
\hline
\end{tabular}
\vspace{1cm}
\smcap{\label{tab-syserror} List of systematic errors contributing
to \protect$ {\fctoD}$, \protect$\fbtoD$
and \protect$\Gcc/\Ghad$. A sign in front of an error indicates 
the direction of change under a positive change of 
the variable.
A detailed explanation of the different errors can be found in the text.}
\end{center}
\end{table}
\begin{itemize}
\item Detector resolution effects: The effects of 
detector resolution modelling
have been discussed in section~\ref{sec-flavsep-sys}.
The total error applicable from these sources to this measurement is 
$\pm 1.8\%$, with the $\dEdx$ error being dominant. 
\item Flavour separation: The systematic errors of the 
flavour separation determined in section~\ref{sec-flavsep-sys} are
used to determine the corresponding systematic errors on the hadronisation
fractions. Many of the errors for $\Gamma_{\ccbar} / \Gamma_{\mathrm had}$ 
are correlated to that of the charm hadronisation fraction. This 
correlation is taken into account in calculating the final errors. 
\item Lepton identification: The systematic errors discussed
in connection with the lepton identification are applicable 
to the measurement of $\fbtoD$. 
\item $\PDstp$ reconstruction: $\Gamma_{\ccbar}/\Gamma_{\mathrm had}$ 
depends on the total rate of $\PDstp$ in $\PZz$ events. All errors
discussed in section~\ref{sec-ndstar} are applicable to this 
measurement as well. 
\end{itemize}
A number of additional errors are introduced through the 
inclusive charm tag:
\begin{itemize}
\item Heavy flavour fragmentation: The efficiency for finding slow pions 
in $\Z0 \to \ccbar$ events 
has been calculated in Monte Carlo.
Systematic errors from the modelling of the heavy flavour fragmentation 
are studied as described in section~\ref{sec-dstar-syserr}.
The total error from this source is
$\pm 1.4\%$ on the slow pion efficiency. 
In table~\ref{tab-syserror} this error is combined 
with the fragmentation error from the b/c separation. 
\item Excited D meson modelling: 
The influence of excited D meson production on 
the slow pion efficiency is studied as described in 
section~\ref{sec-dstar-syserr},
resulting in an efficiency error of $0.1\%$.
\item Jet definition: The resolution of the $p_t^2$ 
signal is dominated by the resolution of the direction 
of the jet axis. Possible modelling problems in the Monte Carlo 
have been checked by comparing the number of charged 
tracks and of neutral clusters used in the calculation of the 
jet axis. The Monte Carlo distributions have been 
re-weighted to the data distributions, and the 
efficiencies are re-determined. An error of $0.9\%$ in 
charm, and $3\%$ in bottom events has been found.
As a cross-check 
the fit has been repeated fixing the fitted 
resolution\footnote{The resolution is defined as the width of
the $p_t^2$ distribution at $50\%$ of its maximum value.} 
of the $p_t^2$ distribution of $0.056\; \GeV$ to its Monte Carlo value of 
$0.061 \;\GeV$. 
Consistent results have been found within the quoted errors.
\item Jet-jet correlation: The efficiency for identifying a slow pion 
is influenced by the presence of an exclusively reconstructed 
$\PDstp$ meson in the other hemisphere. This bias is taken 
into account by calculating the efficiency in Monte Carlo in 
events where a $\PDstp$ meson has been reconstructed in the 
other hemisphere. A small error remains 
if the energy distribution of the secondary jet is 
not modelled properly in the Monte Carlo. This has been 
evaluated by reweighting the Monte Carlo energy distribution to 
that observed in the data, and recalculating the 
efficiency. The resulting error is $0.7\%$ of 
the efficiency.

%
%
%
\item Background subtraction and fitting procedure: 
Systematic effects introduced by the fitting procedure 
for background determination in the double tagged sample 
have been studied in Monte Carlo 
simulation. 
\begin{itemize}
  \item The complete fit has been repeated in Monte Carlo.
  The number of double tags reconstructed is well reproduced within its 
  statistical errors. The difference is 
  $0.5\%$, and this is used an additional systematic error. 
  \item The modelling of the combinatorial background in the fit has 
  been tested in the Monte Carlo by repeating the fit with 
  the true Monte Carlo background. 
  The fit has been done by either constraining the shape 
  of the background to be the same in all double tagged samples, 
  or by fitting it individually in each one. The observed difference 
  of $2.5\%$ has been assigned to this source. 
  \item The shape of the bottom component is estimated from the 
  lepton sample. Possible biases have been investigated in the 
  Monte Carlo by repeating the fit with the true bottom signal. 
  The results agree to within $2\%$, and this has been used as 
  the error. 
  \item The number of fake double tags predicted by the fit in the 
  Monte Carlo has been compared to the known number of fake 
  double tags. They agree to $10\%$. The number of 
  fake double tags in the data has been measured in the fit
  to be $(8.5\pm2.2)\%$ of the combinatorial background. 
  Monte Carlo predicts this  to be $(9.9\pm0.8)\%$, which is compatible 
  within errors  with the measured rate. 
  The error from this source is estimated by varying 
  the fitted fake rate within its error, and by 
  additionally assigning the difference between data and Monte 
  Carlo as an systematic error. In total this gives an 
  error of $3.5\%$ of the signal in charm events, 
  $3.0\%$ in bottom events.
  \item B-mixing: The uncertainty due to mixing in the neutral B sector 
  has been studied by varying the effective mixing parameter
  $\chi_{\rm eff}$ within its error, or $0.5\%$ of the 
  final result. 
  \item Fit function: As a cross check the analysis has been 
  repeated with different parametrisations for both the signal and 
  the background distributions. The background is modelled using 
  a polynomial $(a+b\; p_t^2 + c \;p_t^4)^{-1}$, and no differences 
  are found for the results. The signal function has been replaced
  by a gaussian-like function $\exp(-(a+b\;p_t^2)^2)$, which also 
  results in consistent results. 
\end{itemize}
\item ${\rm g}\to\ccbar$: The contribution to the tagged samples 
from gluon splitting events has been discussed in section~\ref{sec-dstar-syserr}.
It has been re-evaluated for the inclusively tagged jet, where Monte Carlo 
simulation predicts this to be $(0.64\pm0.08)\%$.
A systematic error of 
$0.12\%$ has been assigned to this source.
%
\end{itemize}
A list of all systematic errors  is given in table \ref{tab-syserror}.

\section{Results and Conclusions}
\label{sec-Gcc}
A double tagging technique has been used to measure the 
hadronisation fractions of charm and bottom quarks into 
charged $\PDstpm$ mesons. 
They are determined to be
\begin{eqnarray}
  {\fctoD} &=& 
       \nctoD  \pm \estatctoD \pm \esysctoD \pm 0.004\nonumber\ , \\
  {\fbtoD}&=& 
       \nbtoD  \pm \estatbtoD \pm \esysbtoD \pm 0.003\nonumber\ ,
\end{eqnarray}
where the first error quoted is statistical, the second systematic, 
and the third one due to external branching ratios.

From the number of $\PDstpm$ mesons observed in the 
decay $\PDstp\to\PDz\Pgp, \PDz\to\PK\Pgp$ the multiplicity 
of $\PDstp$ mesons in hadronic Z decays is measured to be 
$$
  \bar n_{\PZz\to\PDstp X} = 
 \nZtoD \pm \estatnZtoD \pm \esysnZtoD  \pm \esysnZtoDBR\ .
$$
Applying bottom tags based on lifetime, jet shape 
and hemisphere charge information in the event the charm and 
the bottom components have been separated, and the individual 
production rates are found to be 
\begin{eqnarray*}
 \Gcc/\Ghad \;{\fctoD}\;\BR(\PDstp\to\PK\Pgp\Pgp) &=& 
  (\nZctoD \pm \estatZctoD \pm \esysZctoD ) \times 10^{-3} \ ,\\
 \Gbb/\Ghad \;{\fbtoD}\;\BR(\PDstp\to\PK\Pgp\Pgp) &=&
  (\nZbtoD \pm \estatZbtoD \pm \esysZbtoD ) \times 10^{-3}\ ,
\end{eqnarray*}
The mean scaled energy of $\PDstp$ mesons in charm events 
is determined from the fragmentation function to be
$$
  \langle x_{\PDstp}\rangle_{\Pc} = \xc \pm \estatxc \pm \esysxc \ .
\nonumber
$$

From the hadronisation fraction $\fctoD$ and the total production 
rate of $\PDstp$ mesons in $\PZz\to\ccbar$ events, 
$\Gamma_{\ccbar}/\Gamma_{\mathrm had}$ is found to be 
$$
\Gcc/\Ghad = \nRc \pm \estatRc \pm \esysRc \pm \ebrRc\ .\nonumber
$$
Here the first error is statistical, the second one describes 
internal systematics, and the last one is due to branching ratio
$\BR(\PDz\to\PK\Pgp)=0.0383\pm0.0012$~\cite{bib-PDG}. The 
correlations between the rate measurement and the 
hadronisation fraction are taken into account in this 
calculation. 
A detailed breakdown of the systematic error is 
given in table~\ref{tab-syserror}. 
The measurements presented in this paper are based on the 
full data sample of almost 4.4 million events collected with the 
OPAL detector at LEP at a centre-of-mass energy of about $91~\GeV$. 
For the first time the charm partial width has been measured 
without significant input from lower energy experiments, 
and in particular without assumptions about the 
centre-of-mass energy dependence of heavy flavour fragmentation. 
In a previous 
OPAL publication the hadronisation fraction $\fctoD$ was
derived from measurements at lower energy ${\mathrm e}^+{\mathrm e}^-$ 
machines to be 
$\fctoD_{\mathrm low\  energy} = 0.262 \pm 0.019 \pm 0.010$, 
where the last error is from the branching ratios $\BR(\PDstp\to\PDz\Pgp)$
and $\BR(\PDz\to\PK\Pgp)$. Assuming that the only correlation 
between the low energy result and the result presented 
in this paper is from the branching ratio $\BR(\PDst\to\PDz\Pgp)$, 
the low energy number is $1.4$ standard deviations higher 
than the OPAL one, which is compatible with the 
assumption that the sources of $\PDstp$ mesons at lower energies 
and at LEP energies are the same. If the low energy hadronisation 
fraction were to be used instead of the one 
measured by OPAL $\Gamma_{\ccbar}/\Gamma_{\mathrm had}$
would be lower by $15\%$.

In conclusion good agreement is found with the 
prediction of the Standard Model~\cite{bib-PDG} of 
\mbox{$\Gamcc/\Gamhad$ = 0.172},
and with other measurements of the same quantities at 
LEP~\cite{bib-DELPHIRc,bib-OPALD*,bib-OPALcharm}. 
\bigskip

\section*{Acknowledgements}
We particularly wish to thank the SL Division for the efficient operation
of the LEP accelerator at all energies
 and for
their continuing close cooperation with
our experimental group.  We thank our colleagues from CEA, DAPNIA/SPP,
CE-Saclay for their efforts over the years on the time-of-flight and trigger
systems which we continue to use.  In addition to the support staff at our own
institutions we are pleased to acknowledge the  \\
Department of Energy, USA, \\
National Science Foundation, USA, \\
Particle Physics and Astronomy Research Council, UK, \\
Natural Sciences and Engineering Research Council, Canada, \\
Israel Science Foundation, administered by the Israel
Academy of Science and Humanities, \\
Minerva Gesellschaft, \\
Benoziyo Center for High Energy Physics,\\
Japanese Ministry of Education, Science and Culture (the
Monbusho) and a grant under the Monbusho International
Science Research Program,\\
German Israeli Bi-national Science Foundation (GIF), \\
Bundesministerium f\"ur Bildung, Wissenschaft,
Forschung und Technologie, Germany, \\
National Research Council of Canada, \\
Hungarian Foundation for Scientific Research, OTKA T-016660, 
T023793 and OTKA F-023259.\\


\newpage
\begin{thebibliography}{99}
\bibitem{bib-ALEPHRb}{ALEPH Collaboration, R.~Barate \etal, {\it A 
measurement of $R_{\Pb}$ using a lifetime-mass tag}, 
CERN PPE/97-017, submitted to Phys.~Lett.~{\bf B};\\
ALEPH Collaboration, R.~Barate \etal, {\it A measurement of 
$R_{\Pb}$ using five mutually exclusive tags},
CERN PPE/97-018, submitted to Phys.~Lett.~{\bf B}.}
\bibitem{bib-DELPHIRb}{DELPHI Collaboration, P.~Abreu \etal,
Z.~Phys. {\bf C70} (1996) 531.}
\bibitem{bib-L3Rb}{L3 Collaboration, O.~Adriani \etal, 
Phys.~Lett. {\bf B307} (1993) 237.}
\bibitem{bib-OPALRb}{OPAL Collaboration, K.~Ackerstaff \etal, 
Z.~Phys. {\bf C74} (1997) 1.}
\bibitem{bib-ALEPHRc}{ALEPH Collaboration, D.~Busculic \etal, 
Z.~Phys. {\bf C62} (1994) 179.}
\bibitem{bib-DELPHIRc}{DELPHI Collaboration, P.~Abreu \etal, 
Phys.~Lett. {\bf B252} (1990) 140;\\
DELPHI Collaboration, P.~Abreu \etal, Phys.~Lett. {\bf B295} (1992) 383. }
\bibitem{bib-OPALD*}
   {OPAL Collaboration, R.~Akers \etal, Z.~Phys. {\bf C67} (1995) 27.}
\bibitem{bib-OPALcharm}{OPAL Collaboration, G.~Alexander \etal, 
Z.~Phys. {\bf C72} (1996) 1.} 
\bibitem{bib-OPALRuds}{OPAL Collaboration, K.~Ackerstaff \etal, 
{\it Measurement of the Branching Fractions and Forward-Backward 
Asymmetries of the $\PZz$ into Light Quarks }, 
CERN PPE/97-63, submitted to Z.~Phys. {\bf C}.}
\bibitem{bib-OPALdetector}{OPAL Collaboration, K.~Ahmet \etal,
  Nucl.~Instr.~Meth. {\bf A305} (1991) 275;\\
  P.P.~Allport \etal, 
  Nucl.~Instr.~Meth. {\bf A324} (1993) 34;\\
  P.P.Allport \etal, 
  Nucl.~Instr.~Meth. {\bf A346} (1994) 476;\\
  O.~Biebel \etal, 
  Nucl.~Instr.~Meth. {\bf A323} (1992) 169;\\
  M.~Hauschild \etal, 
  Nucl.~Instr.~Meth. {\bf A314} (1992) 74.}
\bibitem{bib-OPALmh}{OPAL Collaboration, R.~Akers \etal, Z.~Phys.
  {\bf C65} (1995) 17. }
\bibitem{bib-CONE}
{OPAL Collaboration, R.~Akers \etal, Z.~Phys.~{\bf C63} (1994) 197.}
\bibitem{bib-JETSET}{T.~Sj\"ostrand, Comp.~Phys.~Comm.
  {\bf 82} (1994) 74.}
\bibitem{bib-OPALtune}{OPAL Collaboration, G.~Alexander \etal,
  Z.~Phys. {\bf C69} (1996) 543.}
\bibitem{bib-PETERSON}{C.~Peterson \etal, 
  Phys.~Rev. {\bf D27} (1983) 105.}
\bibitem{bib-LEPEW}{The LEP Collaborations, ALEPH, DELPHI, L3 and OPAL, and 
the LEP Electroweak working group, Nucl.~Instr.~Meth., {\bf A378} (1996) 101;\\
   {The LEP Collaborations, ALEPH, DELPHI, L3 and OPAL, and the 
    LEP Electroweak Working Group, CERN-PPE/96-183, and references
    therein. This paper has been prepared by the LEP 
    collaboration for presentation at major conferences, and contains
    some preliminary numbers.}
}
\bibitem{bib-OPALGOPAL}{J.~Allison \etal,
  Nucl.~Instr.~Meth. {\bf A317} (1991) 47.}
\bibitem{bib-OPALD*AFB}{OPAL Collaboration, G.~Alexander \etal, 
 Z.~Phys. {\bf C73} (1997) 379.}
\bibitem{bib-OPALlAFB}{OPAL Collaboration, G.~Alexander \etal,
Z.~Phys. {\bf C70} (1996) 357.}
\bibitem{bib-OPALANN}{OPAL Collaboration, G.~Alexander \etal, 
Z.~Phys. {\bf C70} (1996) 357.}
\bibitem{bib-OPALleptonID}{OPAL Collaboration, 
             R.~Akers \etal, Z.~Phys. {\bf C60} (1993) 199.}
\bibitem{bib-PDG}{Particle Data Group, 
  Phys.~Rev. {\bf D54} (1996) 1.}
\bibitem{bib-OPALgbb}
   { OPAL Collaboration, R.~Akers \etal , Z.~Phys.~{\bf C65} (1995) 17.}
\bibitem{bib-OPALgcc}
   {OPAL Collaboration, R.~Akers \etal, Phys.~Lett. {\bf B353} (1995) 595.}
\bibitem{bib-DELPHIvcb}{DELPHI Collaboration, P.~Abreu \etal, 
Z.~Phys. {\bf C71} (1996) 539.}
\bibitem{bib-ARIADNE}{
  {L.~L\"onnblad, 
  Comp.~Phys.~Comm. {\bf 71} (1992) 15.}}
\bibitem{bib-Collins}{P.Collins, T.Spiller, J.~Phys. {\bf G11} (1985) 1289.}
\bibitem{bib-Kart}{V.G.~Kartvelishvili, A.K.~Likehoded, V.A.~Petrov, 
Phys.~Lett.~{\bf B78} (1978) 615.}
\bibitem{bib-NASON}{B.~Mele, P.~Nason, Phys.~Lett. {\bf B245} (1990) 635;
                    Nucl.~Phys.~{\bf B361} (1991) 626.}
\bibitem{bib-ALEPHD*}{ALEPH Collaboration, D.~Buskulic \etal ,
Phys.~Lett. {\bf B266} (1991) 218.}
\bibitem{bib-DELPHID*}{
DELPHI Collaboration, P.~Abreu \etal, Z.~Phys. {\bf C59} (1993) 533.}
\bibitem{bib-mix1}{OPAL Collaboration, P.D.~Acton \etal  , 
Z.~Phys. {\bf C67} (1995) 379.}
\end{thebibliography}
\end{document}